\documentclass[iop,apj]{emulateapj}
\slugcomment{Accepted to the Astrophysical Journal}

\usepackage[]{amsmath}
\usepackage[dvipsnames]{xcolor}
\usepackage{bm}
\usepackage{ulem}
\usepackage{rotating}
\usepackage{color}	
\usepackage{mathptmx}
\usepackage{times}
\usepackage[colorlinks=true, citecolor=ao, filecolor=ao, linkcolor=ao, urlcolor=ao]{hyperref}
\usepackage{scrextend}
\usepackage{amsmath}
\usepackage{bm}

\shorttitle{Borg full data analysis}
\shortauthors{Morishita et al.}

\definecolor{midgray}{gray}{0.4}
\definecolor{orange}{rgb}{1,0.5,0}
\definecolor{ao}{rgb}{0.0, 0.0, 1.0}
\deffootnote[0.5em]{0em}{1em}{\textsuperscript{\thefootnotemark}\,}

\usepackage{etoolbox}
\makeatletter
\patchcmd{\NAT@citex}
  {\@citea\NAT@hyper@{\NAT@nmfmt{\NAT@nm}\NAT@date}}
  {\@citea\NAT@nmfmt{\NAT@nm}\NAT@hyper@{\NAT@date}}
  {}
  {}
\patchcmd{\NAT@citex}
  {\@citea\NAT@hyper@{%
     \NAT@nmfmt{\NAT@nm}%
     \hyper@natlinkbreak{\NAT@aysep\NAT@spacechar}{\@citeb\@extra@b@citeb}%
     \NAT@date}}
  {\@citea\NAT@nmfmt{\NAT@nm}%
   \NAT@aysep\NAT@spacechar%
   \NAT@hyper@{\NAT@date}}
  {}
  {}
\patchcmd{\NAT@citex}
  {\@citea\NAT@hyper@{%
     \NAT@nmfmt{\NAT@nm}%
     \hyper@natlinkbreak{\NAT@spacechar\NAT@@open\if*#1*\else#1\NAT@spacechar\fi}%
       {\@citeb\@extra@b@citeb}%
     \NAT@date}}
  {\@citea\NAT@nmfmt{\NAT@nm}%
   \NAT@spacechar\NAT@@open\if*#1*\else#1\NAT@spacechar\fi%
   \NAT@hyper@{\NAT@date}}
  {}
  {}
\makeatother

\newcommand{\myemail}{tmorishita@stsci.edu}
\newcommand{\simgt}{\,\rlap{\lower 3.5 pt \hbox{$\mathchar \sim$}} \raise
1pt \hbox {$>$}\,}
\newcommand{\simlt}{\,\rlap{\lower 3.5 pt \hbox{$\mathchar \sim$}} \raise
1pt \hbox {$<$}\,}
\newcommand{\logm}{\log M_*/\Msun}
\newcommand{\Msun}{{\rm M}_{\odot}}

\newcommand{\kms}{{\rm km~s^{-1}}}

\def\oo{$O_{\rm 350}$} 
\def\jj{$J_{\rm 125}$} 
\def\yy{$Y_{\rm 105}$} 
\def\jh{$JH_{\rm 140}$} 
\def\hh{$H_{\rm 160}$} 

\def\ntot{73374} 

\def\njjcol{15} 
\def\nyycol{19} 

\def\njjvis{5} 
\def\nyyvis{5} 

\def\nyyoap{1}

\def\njjpzori{0}
\def\nyypzori{4}

\def\njjpz{9}
\def\nyypz{5}

\def\njhlv{0}
\def\nyylv{2}

\newcommand{\hst}{{\it HST}}
\newcommand{\jwst}{{\it JWST}}
\newcommand{\spit}{{\it Spitzer}}

\newcommand{\sext}{SExtractor}

\newcommand{\eazy}{{\ttfamily EAzY}}

\newcommand{\borg}{BoRG}
\newcommand{\borgz}{BoRG[z9]}
\newcommand{\as}{\,arcsec}

\newcommand{\affilA}{Space Telescope Science Institute, 3700 San Martin Drive, Baltimore, MD 21218, USA}
\newcommand{\affilB}{School of Physics, University of Melbourne, VIC  3010, Australia}
\newcommand{\affilC}{Department of Physics and Astronomy, UCLA, 430 Portola Plaza, Los Angeles, CA 90095-1547, USA}
\newcommand{\affilD}{Geneva Observatory, University of Geneva, Ch. des Maillettes 51, 1290 Versoix, Switzerland}
\newcommand{\affilE}{ARC Centre of Excellence for All-Sky Astrophysics in 3 Dimensions, Australia}
\newcommand{\affilF}{CASA, Department of Astrophysical and Planetary Science, University of Colorado, Boulder, CO 80309, USA}
\newcommand{\affilG}{Leibniz-Institut fur Astrophysik Potsdam, An der Sternwarte 16, D-14482 Potsdam, Germany}
\newcommand{\affilH}{Department of Physics and Astronomy, 102 Natural Science Building, University of Louisville, Louisville, KY 40292, USA}
\newcommand{\affilI}{Harvard-Smithsonian Center for Astrophysics, 60 Garden St, Cambridge, MA 02138, USA}

\begin{document}
\title{
The bright-end galaxy candidates at $\bm{z\sim9}$ from 79 independent \bm{\hst} fields
}

\author{
T.~Morishita$^{\href{\affilA}{1}}$,
M.~Trenti$^{\href{\affilB}{2,3}}$,
M.~Stiavelli$^{\href{\affilA}{1}}$,
L.~D.~Bradley$^{\href{\affilA}{1}}$,
D.~Coe$^{\href{\affilA}{1}}$,
P.~A.~Oesch$^{\href{\affilD}{4}}$,
C.~A.~Mason$^{\href{\affilC}{5},\href{\affilI}{6},\href{\affilI}{10}}$,
J.~S.~Bridge$^{\href{\affilH}{7}}$,
B.~W.~Holwerda$^{\href{\affilH}{7}}$,
R.~C.~Livermore$^{\href{\affilB}{2,3}}$,
B.~Salmon$^{\href{\affilA}{1}}$,
K.~B.~Schmidt$^{\href{\affilG}{8}}$,
J.~M.~Shull$^{\href{\affilF}{9}}$,
T.~Treu$^{\href{\affilC}{5}}$
}

\affil{$^1$\affilA; \href{mailto:\myemail}{\myemail}}
\affil{$^2$\affilB}
\affil{$^3$\affilE}
\affil{$^4$\affilD}
\affil{$^5$\affilC}
\affil{$^6$\affilI}
\affil{$^7$\affilH}
\affil{$^8$\affilG}
\affil{$^9$\affilF}
\affil{$^{10}$Hubble Fellow}

\begin{abstract}
  We present a full data analysis of the pure-parallel {\it Hubble Space Telescope} (\hst) imaging observations in the Brightest of Reionizing
  Galaxies Survey (\borgz) in Cycle 22.
  The medium-deep exposures with five\,\hst/WFC3IR+UVIS filter bands from 79\,independent sightlines
  ($\sim370$\,arcmin$^2$) provide the least biased determination of number density for $z\simgt 9$ bright galaxies against cosmic
  variance. After a strict two-step selection for candidate galaxies, including dropout color
  and photometric redshift analyses, and revision of previous BoRG candidates, we identify one
  source at $z\sim 10$ and two sources at $z\sim 9$.  The $z\sim10$
  candidate shows evidence of line-of-sight lens magnification
  ($\mu\sim 1.5$), yet it appears surprisingly luminous
  ($M_{UV}\sim-22.6\pm0.3$\,mag), making it one of the
  brightest candidates at $z>8$ known ($\sim0.3$\,mag brighter
  than the $z=8.68$ galaxy EGSY8p7, spectroscopically confirmed by
  Zitrin and collaborators). For $z\sim9$ candidates, we include previous data points at fainter magnitudes and find that the data are well fitted by a Schechter luminosity function with $\alpha=-2.1_{-0.3}^{+0.3}$, $M_{UV}^*=-21.0_{-1.4}^{+0.7}$\,mag, and $\log\phi^*=-4.2_{-0.9}^{+0.6}$\,Mpc$^{-3}$mag$^{-1}$, for the first time without fixing any parameters. The inferred cosmic star formation rate density is consistent with unaccelerated evolution from lower redshift.
\end{abstract}



\section{Introduction}

Cosmic reionization is one of the frontiers of modern astronomy (\citealt{pc1316}; see also \citealt{robertson10}). Of particular interest is which class of astrophysical objects is primarily responsible for this phenomenon, which is likely to happen at $z\sim6$-9 \citep[e.g.,][]{fan00,totani06,shull12,treu13,konno14,schmidt14b,mason18,mason18b}. Determining luminosity functions (LFs) at such an early epoch is therefore a fundamental step in estimating the net contribution from luminous sources, including active galactic nuclei (AGNs) \citep{fan06, treu12, trenti12b, madau14, bouwens15b}.

After a decade of effort, it seems likely that galaxies at the faint end of the LF ($M_{UV}\simgt-18$\,mag) are the key driver in the reionization epoch \citep[][]{ellis13, schenker13, mclure13, oesch13, ono13, atek15, ishigaki15, livermore17, oesch18, ishigaki18}, though with a non-negligible impact from uncertainties, in particular from lens modeling \citep[e.g.,][]{atek18}. 
  
On the other hand, quantifying the number density and properties of the brightest galaxies in the early universe still remains an open yet important field, because these objects constrain the overall shape of the LF \citep[][]{bouwens15, ono18}. Furthermore, these massive objects encode information on in-situ star formation at even earlier times and probe their building blocks \citep[e.g.,][]{stiavelli09}, which are the key to understanding the objects at $z\lesssim7$ \citep[e.g.,][]{ouchi10, sobral15}.

The shape of LFs in the high-$z$/high-luminosity regime is yet poorly sampled and its whole shape is undetermined, partly suffering from low-$z$ interlopers \citep[e.g.,][Bridge, in preparation]{livermore18}. In particular, current observations with the {\it Hubble Space Telescope} (\hst) do not resolve the debate on whether the LF should remain Schechter-like, as it is at lower redshift \citep[e.g.,][]{trenti10, jaacks12, finkelstein15, mason15}, or become a (double) power law with enhancement at the bright end, due to the differential impact of dust attenuation \citep[e.g., ][]{clay15, finkelstein15}, AGN feedback \citep[][]{finlator11}, and/or mass quenching \citep{birnboim03, peng10, woo13}.

Legacy field investigations with \hst\ have covered a substantial area ($\simgt 800$\,arcmin$^2$ down to $\sim27$\,mag; \citealt{grogin11, koekemoer11}), and the latest samples are approaching $\sim1000$ sources \citep[e.g.,][]{bouwens15}. However, at the bright end one concern in addition to the paucity of candidates is the strong clustering of these sources, with bias factor exceeding $b>8$ \citep{baronenugent14}. This in turn implies that contiguous fields can be significantly affected by cosmic variance \citep[][]{trenti08}. 

The Brightest of the Reionizing Galaxies (\borg) Hubble Space Telescope survey (\citealt{bradley12,trenti12}) has
been designed specifically to contribute towards an unbiased
measurement of the number density of the brightest and most clustered
sources at $z\gtrsim 8$, by taking advantage of pure-parallel (random
pointing) opportunities to cover a wide area with medium-deep
optical and IR imaging ($5\,\sigma$ limiting magnitude
$m_{AB}\sim 26.5$), over more than 100 {\it independent} sightlines so far. 

The initial \borg\ survey aimed at the bright end of the LF at $z\simgt8$, using four filters on WFC3 \citep[][]{trenti11, trenti12, bradley12, bernard16}.
\citet{schmidt14b} explored 350\,arcmin$^2$ and found 38 $Y$-band dropout candidates with $L\approx L_*$,
providing one of the strongest constraints on the shape of the $z \sim 8$ LF.
A continuation of the survey, \borgz\ (GO13767, PI. M. Trenti),
which we present in this study, is optimized for galaxies at higher redshift ($z\simgt 9$) with an updated set of five WFC3IR/UVIS filters. 
Preliminary results from this survey were presented in \citet[][hereafter C16]{calvi16} by using $\sim1/3$ of the full data, demonstrating the strength of the new strategy.

In this paper, we present the full data set of \borgz---a collection of 79 independent sightlines ($\sim370$\,arcmin$^2$) down to $\sim27$\,mag---which is the least biased survey in the early universe in terms of cosmic variance.\footnote{{For example, cosmic variance for our 79\,independent FoVs reduces to $\sim26\%$ of the value for a single contiguous survey with the same area coverage \citep{trenti08}.}
}
We base our analysis on the well-studied color selection method presented in C16. 
We then further constrain the color-selected candidates with additional criteria, which include photometric redshift priors based on a theoretical model as well as follow-up \spit\ photometry.
Our new selection also updates the candidates found in C16, providing a
final set of one candidate at $z\sim10$ and two at $z\sim9$, respectively, from all \borgz\ fields.
Finally, taking into account simulations of source recovery and completeness, and with the inclusion of data for faint galaxies from the literature, we determine the LF at $z\sim9$ for the first time without fixing any parameters.

Throughout the paper, magnitudes are quoted in the AB system
\citep{oke83}. We assume $\Omega_m=0.3$, $\Omega_\Lambda=0.7$,
$H_0=70\,\kms\, {\rm Mpc}^{-1}$.  We refer to the WFC3 filters
F350LP, F105W, F125W, F140W, and F160W as \oo\ (optical), \yy, \jj, \jh,
and \hh, respectively.


\section{Data and Sample Selection}\label{sec:data}

In the present study, we use the full data set of \borgz, which
consists of 79 independent field of views (FoVs) $\sim370$\,arcmin$^2$). These data
have been made publicly available as Version 3 of the BoRG survey as
fully calibrated and aligned science images.\footnote{\url{https://archive.stsci.edu/prepds/borg/}}
This analysis includes the data previously studied in C16, which consists of 28 initial FoVs ($\sim130$\,arcmin$^2$).
Inclusion of the previous data set allows us to check the consistency and to see if updates made in the present analysis affect the final results.\footnote{Since we update the source detection parameters from C16, target IDs between the two studies are not identical.}

\subsection{Survey Design}\label{ssec:design}
\borgz\ is a pure-parallel \hst\ imaging survey with five broadband filters spanning from the NUV/optical band of WFC3/UVIS (\oo) to near-IR (\yy/\jj/\jh/\hh) of WFC3/IR. Its medium deep exposures ($\sim2$-$5$\,ks) in random sightlines optimize for detecting bright galaxies, and reach typical limiting magnitudes of $m_{5\,\sigma} \sim27$\,mag measured in an $0.\!\arcsec32$ radius aperture (see Appendix~\ref{sec:Aa}).

The major update in \borgz\ from our previous studies \citep[HST Cycles 17 and 19;][]{trenti11,trenti12,bradley12,schmidt14b} is the use of a long-pass filter in the optical wavelength range. \oo\ covers $\sim0.3\,\mu$m to 1.0\,$\mu$m, while many other studies use a single or multiple broadband filters in the optical range.
The choice optimizes rejection of contaminants for the selection of $z>8$ sources. 

For our primary science goal, the collection of \jj\hh\ and \yy\jh-dropout sources, persistence is of particular concern because of the possibility of introducing an artificial coherent signal into the near-IR bands.
In each visit of this program, we therefore arranged the sequence of WFC3/IR filters to minimize the impact. 
As detector persistence decays over time, with approximate power-law behavior, any saturated target observed in a previous visit most affects the initial part of the pure-parallel orbit. 
The general strategy therefore is to observe in the dropout filter as early as possible in the orbit, ideally placing the \yy-band first.
However, doing so would result in increased time-varying backgrounds from scattered-light Earth glow from helium in the upper atmosphere \citep[caused by a He $10830$\,\AA\ line;][]{brammer14}.
Because the intensity of the scattered Earth glow decreases 
with increasing target angles above the bright Earth limb, we opted 
to place the \yy\ observations as the second filter after \jj\ (or sometimes 
third when the orbit begins with \oo) in an orbit.  
Whenever possible, observations in \jh\ and \hh\ follow those filters. 
With this strategy, persistence features are essentially prevented
from contaminating the \jj\hh\ and \yy\jh-dropout selections.

To ensure good sampling of the IR exposures, we opted for reading every 50\,s (SPARS50).  
While the majority of cosmic rays are rejected by the calibration pipeline, owing to the multiple non-destructive readouts of the WFC3IR detector, a small fraction may survive in the calibrated image. 
We thus split the total integration in each IR filter into at least two individual exposures.  
\oo\ exposures obtained with the UVIS CCD are split into more than two sequences (each 450-900\,seconds) for optimal cosmic-ray rejection. 
Our design choices are aimed at maximizing the data quality, though a
small price is paid in signal-to-noise ratio (S/N) by increasing the number
of individual exposures.

\subsection{Data Reduction}\label{ssec:reduction}
We reduced the raw imaging data by using the \hst\ pipeline in the standard manner.
In addition to running calwf3, processing of \oo\ included a correction for the Charge-Transfer Efficiency effect \citep[CTE19;][]{noeske12, anderson14}.
For all filters,  we also performed a customized extra step to remove
residual cosmic rays and/or detector artifacts, such as unflagged hot
pixels, by using a Laplacian edge filtering algorithm
\citep[LACOSMIC;][]{vandokkum01} that was used for the previous BoRG
data reduction \citep[][see also C16]{bradley12, schmidt14b}.

The pixel scale is set to 0.08\as\,/pixel with pixfrac to 0.75, as in C16.
Since most of our images are taken without dithering, we use a slightly larger pixel scale than typical non-parallel observations ($\sim0.06$\,\as/pixel).
The only change from C16 is made in one of the parameters of cosmic-ray detection, where we set cr\_threshold to $3\,\sigma$ and $1.5\,\sigma$ (compare $3.9\,\sigma$ and $5\,\sigma$ in C16) for UVIS and IR detectors, respectively, in the LACOSMIC Python package\footnote{\url{http://lacosmic.readthedocs.io/en/latest/}}.
This significantly reduces the residual cosmic rays compared to the previous data products.

This reduction process generates science and rms maps for the five filter bands, in addition to a combined \jh+\hh\ map for the use of detection images (see below).
It is noted that the RMS maps generated by the pipeline have arbitrary infinity values for, e.g., dead pixels and/or those with a cosmic-ray flag.
Such an artificial value in fact affected the previous sample selection in C16 (see Section~\ref{sssec:jj}).
{We replace those values with 0, so that a detection algorithm \citep[i.e. SExtractor;][]{bertin96} ignores them; otherwise it would falsely return infinity values for fluxes in such regions.}

The RMS map of each filter band generated by the pipeline is then scaled so that it represents the true uncertainty, including correlated noise \citep[][]{casertano00}.
To account for this, we follow the scaling method presented by \citet{trenti11}.
Briefly, we measure the median sky in the empty region in each mosaic image by using \sext\ \citep{bertin96} with the same setup parameters as the source detection.
Each RMS map is then scaled so that {the dispersion in flux of the sky region ($f_{aper}$, in an aperture of $r=0.\!\arcsec32$) and its estimated error ($e_{aper}$) are consistent.}
The median values of the scale factor are 1.174, 1.195, 1.209, 1.233, 1.708 for \hh, \jh, \jj, \yy, and \oo, respectively, which are consistent with C16.
This process also returns limiting magnitudes (i.e. median of $f_{aper}$ in each field), which are used for upper limits of non-detection.

Ten out of the 89\,original survey fields are discarded from our final analysis, because they failed in the acquisition of a guide star or are dominated by a number of bright stars.
The list of fields is summarized in Appendix~\ref{sec:Aa}.

The point-spread functions (PSFs) of different filter images are not matched in this study, because it is challenging to accurately measure the PSF in each of our data with relatively shallow exposures. However, as the apparent size of our target galaxies is typically small, in addition to the use of optimized photometry (i.e. isophotal flux; see Section~\ref{ssec:cs}), the effect is minimal \citep[see also][]{oesch07, trenti12, holwerda14b}.

\begin{figure*}
\centering
\includegraphics[width=0.85\textwidth]{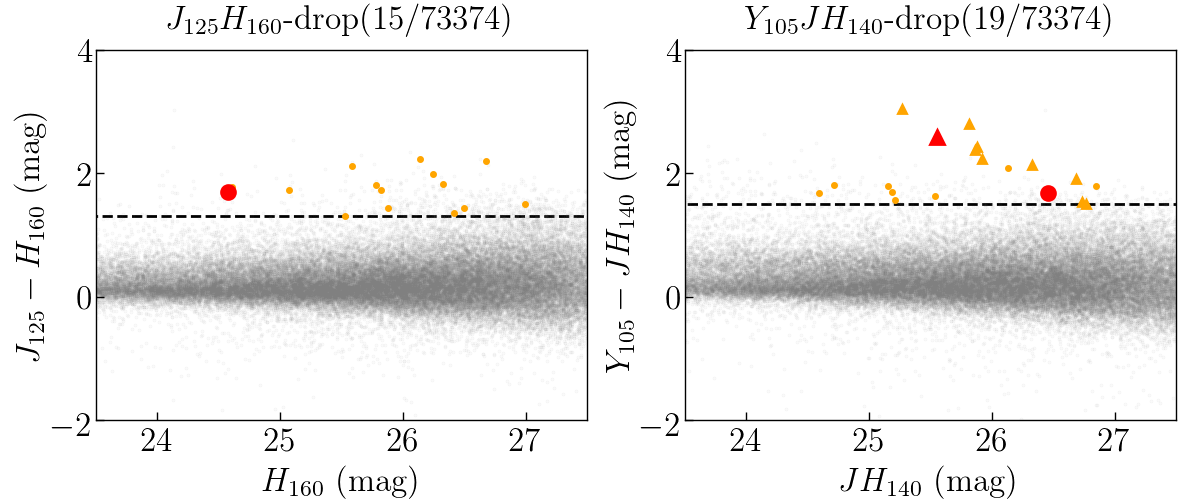}
\caption{
Color-magnitude diagram for the 79 \borgz\ fields.
Out of \ntot\,detected sources (gray points), \njjcol\ and \nyycol\ objects are selected as initial $z\sim10$ (left) and $z\sim9$ (right) candidates, respectively (orange circles for those with measured color and triangles for those with a lower limit). The criterion for the Lyman-break color cut is shown with a dashed line.
We apply visual inspection (Section~\ref{ssec:vis}), photometric redshift selection (Section~\ref{ssec:pz}), and \spit\ photometric selection (Section~\ref{ssec:spt}) to the initial candidates in order to select the final candidates (large red symbols).
}
\label{fig:drop}
\end{figure*}

\subsection{Detection and Color Selection of Dropout Galaxies}\label{ssec:cs}

We first detect galaxies using \sext\ in the \jh+\hh\ stacked image. The detection parameters are set as C16---{DETECT\_MINAREA}\,$=9$, {NTHRESH}\,$=0.7\,\sigma$, {DEBLEND\_NTHRESH}\,$=32$, {DEBLEND\_MINCONT}\,$=0.01$. The exception is the convolution size (FWHM of Gaussian) for detection in \sext, which is changed from 2\,pixels to 5\,pixels. This reduces false detection of, e.g., discrete noise at the edge of the detector and residual cosmic rays.
The photometry is performed in the dual-imaging mode, based on the \jh+\hh\ detection.
{With this setup, we detect \ntot\ objects from all BoRG fields in this study.}

We then select those with $S/N_{JH_{140}+H_{160}}>8$ for robust detection, and with CLASS\_STAR $<0.95$ to avoid stars.
Signal-to-noise ratios are calculated from the measurement and error of the isophotal flux of \sext.
Compared to other schemes of photometry, the choice flexibly corresponds to source morphology.
For example, fixed-aperture photometry tends to underestimate the S/N in \oo, which would mistakenly select low-$z$ galaxies as the dropout candidates.

To minimize the contamination by low-$z$ interlopers, we also limit half-light radius along the major axis of \sext\ to $0.\!\arcsec3$ (or $\sim1.7$\,kpc at $z\sim6$), the observed upper limit for $z>6$ galaxies at the present magnitude limit \citep[e.g.,][]{oesch10,shibuya15,holwerda15,kawamata17}.

We then follow the color-cut criteria proposed by C16 for the selection of $z\sim10$ and $z\sim9$ candidates, whose $95\%$ confidence regions correspond to $7.7<z<9.7$ (with a peak at $z\sim8.7$) and $z>9.3$, respectively;

\

\

\begin{itemize}
\item $z\sim10$ candidates (\jj\hh-dropouts)
$$S/N_{350}<1.5$$
$$S/N_{105}<1.5$$ 
$$S/N_{160}\ge6$$
$$J_{125}-H_{160}>1.3$$

\item $z\sim9$ candidates (\yy\jh-dropouts)
$$S/N_{350}<1.5$$ 
$$S/N_{140}\ge6$$  
$$S/N_{160}\ge4$$
$$Y_{105}-JH_{140}>1.5$$
$$Y_{105}-JH_{140}>5.33\cdot(JH_{140}-H_{160})+0.7$$
$$JH_{140}-H_{160}<0.3$$ 
\end{itemize}

Colors are calculated after correcting for Galactic dust extinction, where $E(B-V)$ is retrieved from the NASA/IPAC infrared archive \citep{schlegel98,schlafly11}.\footnote{\url{http://irsa.ipac.caltech.edu/applications/DUST/}}
We assume the canonical Milky Way dust law \citep{cardelli89} to calculate the extinction in each filter.

The observed flux is scaled by $C_{160}=f_{\rm AUTO, 160}/f_{\rm iso, 160}$ for the following analysis, where $f_{\rm AUTO, 160}$ and $f_{\rm iso, 160}$ are the AUTO (i.e. total) flux and isophotal flux of \sext\ measured in \hh, so as to correct the flux missed in the isophotal flux. 
The scale factor of individual objects is applied to the other four bands to uniformly scale the fluxes.

The selected sources are shown in Figure~\ref{fig:drop}.
There are \njjcol\ and \nyycol\ sources selected with the color criteria for $z\sim10$ and $z\sim9$ candidates, respectively, up to $H_{160}\sim27$\,mag.
The color near the Lyman break (\jj\hh\ and \yy\jh) is calculated with the 1\,$\sigma$ limiting magnitude of each image when the magnitude at the shorter wavelength has $S/N< 1$, and shown as a lower limit in Figure~\ref{fig:drop}.

While our color selection mostly excludes sources with optical detection, we found in the visual inspection (Section~\ref{ssec:vis}) that some candidates show tiny blobs in \oo, despite satisfying $S/N_{350}<1.5$ in the isophotal magnitude.
Given its bright/extended appearance in longer wavelength filters, we consider this as sub-galactic scale/patchy star formation from dusty galaxies at lower redshifts (see Section~\ref{ssec:pz}).
We therefore exclude those with $S/N_{\rm 350, ap.}>1.5$ in addition to the criteria above, where $S/N_{\rm 350, ap.}$ is a signal-to-noise ratio measured in a small aperture (radius of 2\,pixels, or 0.\!\arcsec16).
The aperture photometry maximizes the S/N inside objects for this case, while isophotal flux is typically measured in a larger aperture from detection images (\jh+\hh\ band in this study) and could lower S/Ns at shorter wavelength.
This excludes one of the $z\sim9$ candidates selected with the default criteria, while there are no such objects in $z\sim10$ candidates.
The remaining candidates are taken through the following selection process.

\subsection{Visual Inspection}\label{ssec:vis}
Four of the authors (T.M., M.T., M.S., and R.C.L.) visually checked the color-selected candidates, because we are not able to exclude artificial objects, such as PSF spikes and residual cosmic rays with only the color criteria. We exclude five objects selected as $z\sim10$ candidates and five selected as $z\sim9$ candidates. Those are stellar spikes, cosmic rays, or inappropriately deblended fragmentation of large galaxies.

We also check the persistence among candidates by inspecting the individual fields.\footnote{\url{https://archive.stsci.edu/prepds/persist/search.php}}
We conclude that none of the remaining candidates is affected by persistence.

\subsection{Photometric Redshift Selection}\label{ssec:pz}

\subsubsection{A flat prior}\label{sssec:fp}
While the color selection is known as a standard method to efficiently select high-$z$ candidates with minimum filter requirements, the result is binary (i.e. yes or no). {In other words, the color selection by itself does not tell us how reliable those selected candidates are.}
To quantify the candidate selection, we use photometric redshifts as additional information.
While their accuracy strongly depends on the number of filters and their depth, photometric redshifts allow us to investigate the consistency of the color selection, quantify the robustness, and eliminate unlikely candidates, if any.

We use \eazy\ \citep[][version of 2015 April 20]{brammer08} to derive photometric redshifts for our candidates selected in Section~\ref{ssec:cs}.
The default template set (v1.3) is used with a minor but key
modification that imposes the template age to be lower
than the age of the universe at the redshift derived.
The choice of the template set is based on the test by \citet{brinchmann17}, where they tested all template sets in \eazy\ and found that the least biased template for high-redshift galaxies is the v1.3 template.
This template is based on the original \eazy\ template that was developed from synthetic galaxy photometry using SAMs, but also includes several additional spectral energy distributions (SEDs) \citep{bruzual03, maraston05, erb10} and emission lines \citep{ilbert09}.
{We also add young star forming galaxy templates with dust attenuation up to $A_V=4.5$\,mag, because those at $z\sim2$ resemble SEDs of high-$z$ galaxies \citep{salmon18}.}

While there are several choices of priors for low-$z$ galaxies (e.g., as a function of apparent magnitude/color), we start with a flat prior (but see also below).
We adopt the redshift at the peak probability distribution as the best-fit redshift ($z_{a}$).

With this setup, we found an excellent agreement for all of $z\sim10$ candidates with the target redshift of the color selection, while this decreases to $\sim60\%$ for $z\sim9$ candidates.
Those with inconsistent photometric redshifts show widely extended probability distributions.
This is partly because the color near the Lyman break is not as strong for those galaxies as for those of other candidates with consistent redshifts.
The color of the Lyman break for the strongest candidates is
$\simgt2$\,mag, while $\sim50\%$ of the rejected sources have a bluer
color. 
That is, they are located near the boundary of the selection box and
could have been scattered into the box due to photometric noise.
We here exclude four of $z\sim9$ candidates that have a low-$z$ peak probability of $>20\%$ of the
high-$z$ peak probability.

\begin{figure}
\centering
\includegraphics[width=0.49\textwidth]{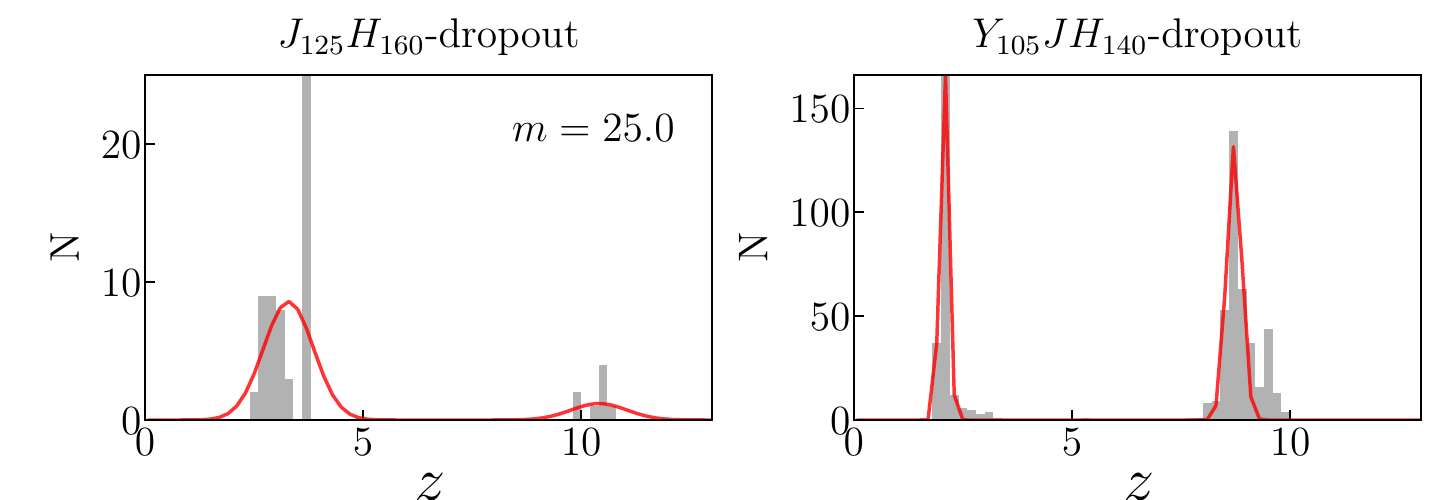}
\includegraphics[width=0.49\textwidth]{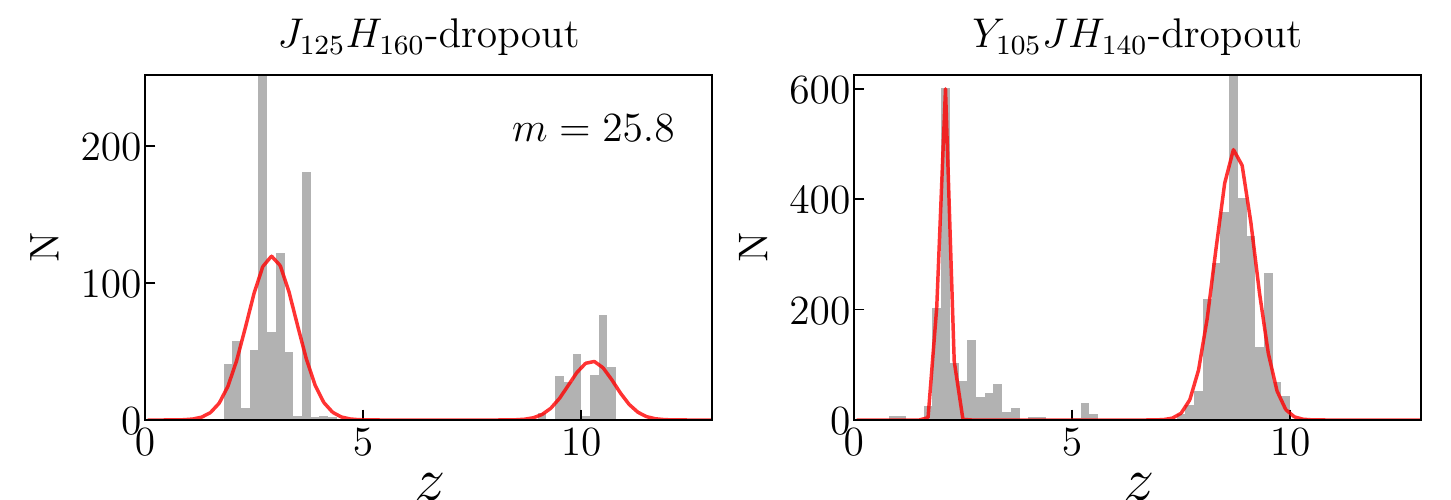}
\includegraphics[width=0.49\textwidth]{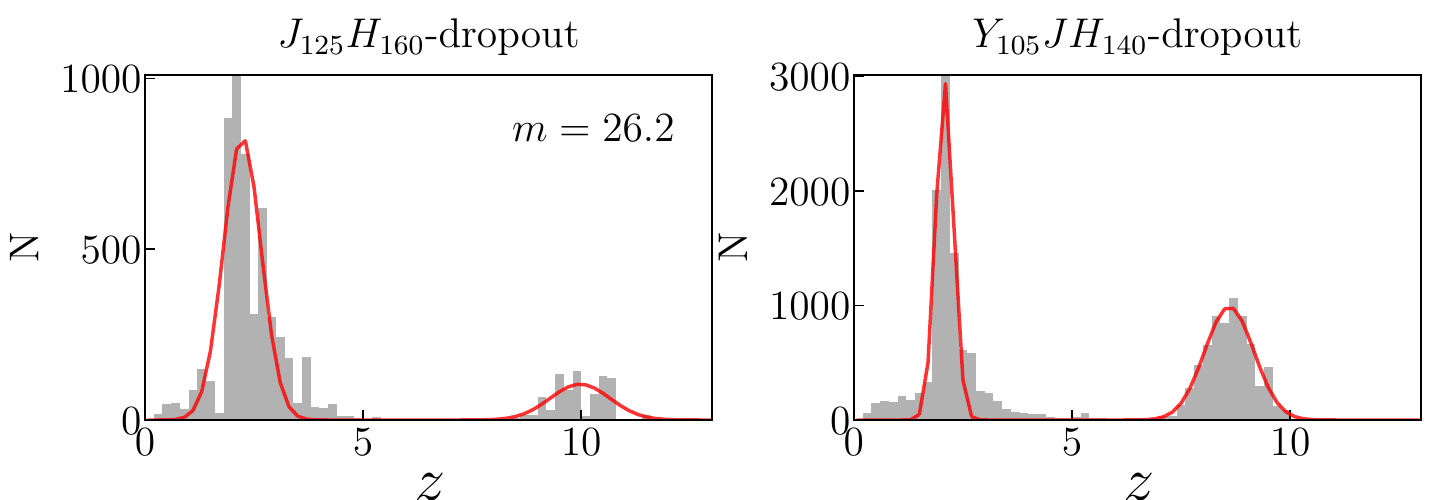}
\includegraphics[width=0.49\textwidth]{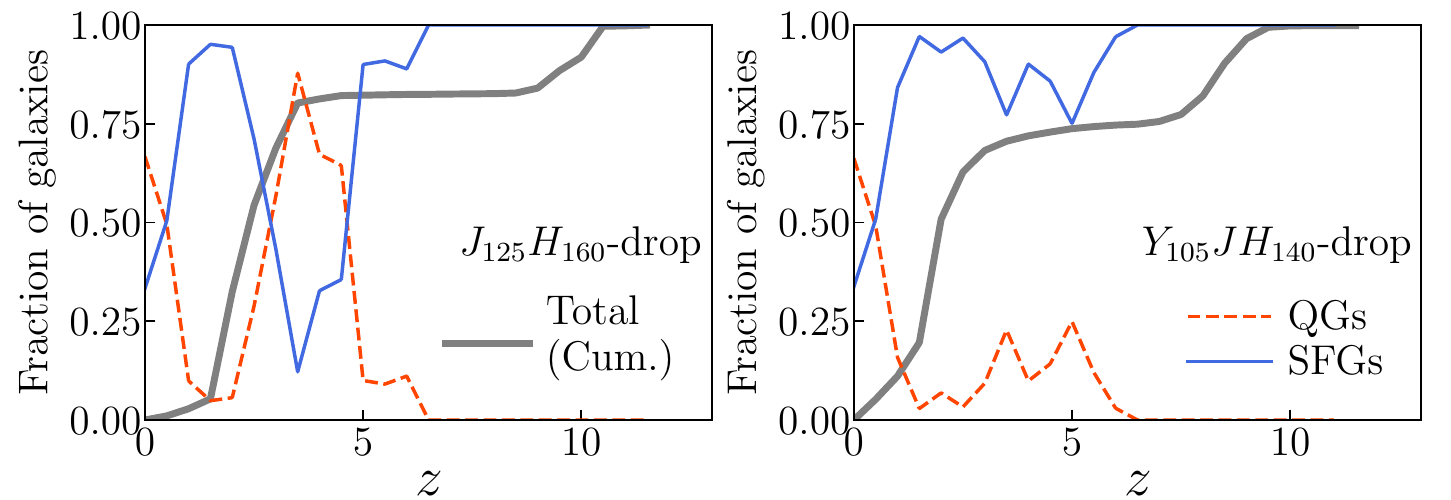}
\caption{
Examples of intrinsic redshift distributions of dropout candidates selected in the mock catalog at different magnitude bins.
Each histogram is fitted with a double-peaked Gaussian (red solid lines) and used as a photometric redshift prior (Section~\ref{sssec:ap}).
The best-fit parameters for Gaussian fitting are summarized in Table~\ref{tab:gauss}.
The bottom panel shows the fraction of star-forming (blue solid lines) and quiescent galaxies (red dashed lines) selected as dropout candidates ($H_{160}>24.5$\,mag).
The total number of mock galaxies is shown as a cumulative fraction (gray solid lines).
While low-$z$ interlopers of $z\sim10$ candidates consist of both populations, those of $z\sim9$ candidates are mostly star-forming galaxies.
}
\label{fig:sim}
\end{figure}

\begin{deluxetable}{cccccccc}
\tabletypesize{\footnotesize}
\tablecolumns{8}
\tablewidth{0pt} 
\tablecaption{Gaussian parameters for redshift priors}
\tablehead{\colhead{$m_{160}$} & \colhead{$A_{z,1}$} & \colhead{$\mu_{ z,1}$} & \colhead{$\sigma_{z,1}$} & \colhead{$A_{z,2}$} & \colhead{$\mu_{z,2}$} & \colhead{$\sigma_{z,2}$} & \colhead{$f_{\rm int.}^{\ \dagger}$}
}
\startdata
\cutinhead{\jj\hh-dropouts}
$25.00$ & $0.87$ & $3.29$ & $0.57$ & $0.13$ & $10.43$ & $0.58$ & $0.83$\\
$25.25$ & $0.66$ & $3.13$ & $0.60$ & $0.34$ & $10.53$ & $0.12$ & $0.86$\\
$25.50$ & $0.57$ & $3.13$ & $0.62$ & $0.43$ & $10.51$ & $0.12$ & $0.82$\\
$25.75$ & $0.74$ & $2.90$ & $0.56$ & $0.26$ & $10.24$ & $0.52$ & $0.76$\\
$26.00$ & $0.77$ & $2.51$ & $0.61$ & $0.23$ & $10.11$ & $0.62$ & $0.77$\\
$26.25$ & $0.89$ & $2.23$ & $0.43$ & $0.11$ & $9.97$ & $0.69$ & $0.86$\\

\cutinhead{\yy\jh-dropouts}
$24.50$ & $0.43$ & $2.12$ & $0.08$ & $0.57$ & $8.76$ & $0.14$ & $0.30$\\
$24.75$ & $0.56$ & $2.09$ & $0.10$ & $0.44$ & $8.72$ & $0.16$ & $0.37$\\
$25.00$ & $0.57$ & $2.07$ & $0.10$ & $0.43$ & $8.72$ & $0.17$ & $0.38$\\
$25.25$ & $0.61$ & $2.06$ & $0.09$ & $0.39$ & $8.75$ & $0.26$ & $0.36$\\
$25.50$ & $0.70$ & $2.16$ & $0.06$ & $0.30$ & $8.76$ & $0.34$ & $0.34$\\
$25.75$ & $0.55$ & $2.08$ & $0.12$ & $0.45$ & $8.74$ & $0.46$ & $0.35$\\
$26.00$ & $0.60$ & $2.08$ & $0.16$ & $0.40$ & $8.70$ & $0.51$ & $0.44$\\
$26.25$ & $0.75$ & $2.08$ & $0.20$ & $0.25$ & $8.61$ & $0.57$ & $0.65$\\
$26.50$ & $0.86$ & $2.07$ & $0.28$ & $0.14$ & $8.47$ & $0.66$ & $0.84$\\
$26.75$ & $0.88$ & $1.99$ & $0.59$ & $0.12$ & $8.31$ & $0.78$ & $0.92$\\

\enddata
\tablecomments{
For \jj\hh-dropout candidates with $m_{160}<25$ and $>26.25$\,mag, reliable Gaussian fitting parameters are not available from the small sample size. 
We instead adopt those from $m_{160}=25$ and 26.25\,mag for candidates, respectively.\\
$\dagger$: Fraction of low-$z$ interlopers.
}
\label{tab:gauss}
\end{deluxetable}


\subsubsection{Building a practical prior}\label{sssec:mp}
A flat prior is often assumed for Lyman break galaxies at high $z$,
since it is still unclear weather applying priors from low redshift is
appropriate \citep[][]{benitez00, salmon17, salvato18}. On the other hand, we
{\it know} that the LF of galaxies evolves strongly
from $z\sim10$ to 0 \citep[e.g.,][]{behroozi13, bouwens15, mason15b,
  kelson16,oesch18}. Combining the intrinsic evolution with changes in
the distance modulus with redshift, we should, at a given luminosity,
expect many more galaxies at lower redshift in a given survey area \citep[][]{stiavelli09,vulcani17}. This
effect has not been taken into account yet in this study.

Motivated by this fact, we design an empirical prior based on a
combination of deep observations and theoretical modeling, which aims
at quantifying the relative abundance of high-$z$ sources versus
low-$z$ interlopers for color selected samples using the dropout
technique. To construct such prior, we start from a mock galaxy catalog by 
\citet[][]{williams18}. While the catalog (v1.0) has been primarily aimed at
planning observations with the {\it James Webb Space Telescope (JWST)},
based on real deep surveys with multi-band including most \hst\ broadband filters, it
reproduces observed galaxy properties self-consistently, such as LF evolution up to $z\sim8$. Therefore, it is well suited
for our purposes.

The model flux in the catalog for each galaxy is perturbed to simulate observed fluxes. 
We assume Gaussian noise based on our
limiting magnitudes. 
In addition to this
random noise added to the model flux, we fluctuate the intrinsic flux
by $0.3$\,dex at each redshift bin (whose width is $\delta z=0.1$), to partially take
into account the effect of cosmic variance
\citep[][]{somerville04, trenti08}, which is not reflected in the
original catalog based on relatively small-volume observations.  
{While this is a simplified approach, shifting magnitude artificially adds $\sim0.4$\,dex variation in galaxy number densities at $M_{UV}\simlt-20$.}
We repeat this 10000 times (i.e. 10000 mock catalogs with
different cosmic fluctuations, where each object has Gaussian random
noise added to its flux), then select dropout candidates with our color
criteria from all catalogs.  
The only difference is the use of F435/606/814W bands in the mock catalog, because
\oo\ flux is not listed in the original catalog.
We estimate the weighted mean flux from the three optical bands as
\begin{equation}
f_{\rm 350} = \sum_{i} f_i \times w_i / \sum_{i} w_i
\end{equation}
where the weight is the convolution of transmission in each filter with \oo,
\begin{equation}
w_i = T_i \ast T_{\rm 350}.
\end{equation}

From the mock catalog, we estimate the fraction of low-$z$ interlopers that are selected with
our color selection, as a function of magnitude and redshift.
We define low-redshift interlopers as those with intrinsic redshift at $z<8$ but satisfy the color-color selection criteria in Section~\ref{ssec:cs}.

Figure~\ref{fig:sim} shows examples --- the {\it intrinsic} redshift distributions of the dropout candidates at a given {\it observed} \hh\ magnitude. The fractions of interlopers ranges from $\sim70\%$ ($\sim40\%$) to $\simgt85\%$ ($\sim90\%$) for $z\sim10$ ($z\sim9$) candidates from \hh$\sim24.8$\,mag to $26.5$\,mag, which is consistent with the analysis of \citet{vulcani17}.  

The interloper population is dominated by $z\sim2$-3 galaxies. The fraction of
star-forming/quiescent populations in the low-$z$ interlopers is shown
in the bottom panel of Figure~\ref{fig:sim}. While
the low-$z$ interlopers of $z\sim10$ candidates consist of both
populations, i.e. the star forming population at $z\sim2$ and quiescent
population at $z\sim3.5$, those of $z\sim9$ candidates are
dominated by the star-forming population at $z\sim3$.
For \yy\jh\ dropout candidates, the fraction of low-$z$ interlopers increases as it approaches the magnitude (\hh$\sim26.5$), because of large photometric flux errors start to affect the sample selection.
{On the other hand, for \jj\hh\ dropout candidates it shows a rather flat distribution over the magnitude range. 
This is partly due to different selection effects for the color selection criteria---while \yy\jh\ dropout candidates are mainly affected by photometric error, \jj\hh\ dropout candidates are contaminated mainly by systematic effects due to, e.g., a smaller number of filters to characterize the Lyman break.}

\subsubsection{Application of the redshift prior}\label{sssec:ap}

{In general, a redshift posterior is calculated as
\begin{equation}\label{eq:post}
p(z | C,m) \propto p(z | m) p(C | z),
\end{equation}
where $p(C | z)$ is a likelihood (derived by \eazy) given the data $C$.
$p(z | m)$ is a prior as a function of $m$ \citep[e.g., magnitude;][]{benitez00}.
In the present study, since we first select galaxies with color criteria, an additional term is implicitly implemented in Equation~\ref{eq:post}:
\[
    p(C) = 
\begin{cases}
    1,& \text{if objects satisfy the color selection}\\
    0,              & \text{otherwise}
\end{cases}
\]
With this, redshift distributions derived in the previous section (Fig.~\ref{fig:sim}) can be used as priors, $p(C) p(z|m)$, in the calculation of posterior for our preselected candidates.
}

To build a functional form for priors, we fit the
redshift distribution with a dual-peak Gaussian model,
$p(C) p(z| m)=\sum_{i=1,2} G_i(z| I_i, m)$, where
$I_{i}=\{A_{z,i}, \mu_{z,i}, \sigma_{z,i}\}$ is the parameter set for
the Gaussian fit,
$G_i(z| I_i)=A_{z,i}~{\rm exp}\big[-(z-\mu_{z,i})^2/2\sigma_{z,i}^2\big]$.
The best-fit parameters ($\chi^2$ minimization) are summarized in
Table~\ref{tab:gauss}.  Due to the small number of galaxies, Gaussian fits
for $z\sim10$ candidates in the $H_{160}\simlt25$ and $\simgt26.3$\,mag bins are not
available.  We instead extrapolate without changing the parameter values
from the 25\,mag and 26.25\,mag bins to those magnitude bins,
respectively.

Given the smaller error in \hh\ magnitudes ($<0.2$\,mag) compared to the
magnitude bin of the prior, we simply
multiply the fitted Gaussian function by each redshift probability
distribution derived with a flat prior in the earlier part of this
section.  We use the redshift at the peak of the posterior distribution as the final redshift ($z_{prior}$).  We
exclude those candidates with a low-$z$ peak probability of $>20\%$ of the
high-$z$ peak probability, based on the typical contamination
fraction at $z\sim10$ \citep{pirzkal13}.  Thus, the photo-$z$
selection assures $80\%$ reliability, or equivalently a contamination
fraction $f_{\rm cont}=0.2$ (Section~\ref{ssec:nd}). 
We reject nine and five of the remaining $z\sim10$ and $z\sim9$ candidates, respectively.

While the mock catalog of \citet{williams18} reproduces the observed
LFs at $z\simlt8$ sufficiently well, we note that galaxies at $z>8$,
especially at the bright end, are yet uncertain, and a priori
assumption, such as the Schechter-form LF, may no longer be
appropriate.  It is also possible that samples might be somewhat
affected by currently unaccounted for contamination by a class of rare
intermediate-redshift galaxies with observed colors similar to those
of $z\gtrsim9$ sources.  {Yet, all our final candidates have a low
  likelihood of being interlopers (less than $20\%$), regardless of
  the application of the prior, strengthening confidence in their
  high-$z$ nature.  In fact, the application of the prior suppresses
  high-redshift probabilities for most of our cases
  (Table~\ref{tab:gauss}), and thus it offers a more conservative
  candidate selection.}  Of course, further follow-up of our
candidates with deeper imaging, or spectroscopy, would represent an
independent assessment of the selection.

\subsection{Spitzer/IRAC photometry}\label{ssec:spt}

For further validation of our candidates, we check the availability of
public \spit\ data\footnote{\url{http://archive.spitzer.caltech.edu}}
acquired from multiple \spit/IRAC $3.6\,\mu{\rm m}$ observations of BoRG fields (PIs
S. Bernard, R. Bouwens, B. Holwerda; $m_{5\sigma}\sim25$-$26.5$\,mag).  While the data set does not offer
complete coverage of the entire \borgz\ survey, IRAC photometry is capable of efficiently excluding low-$z$ interlopers in the case of significant excess of flux compared to the \hh\ band such as, e.g., quiescent+dusty SEDs at $z\sim2.5$ \citep{holwerda15, oesch16,salmon18}. 

The $z\sim10$ candidate has an IRAC $3.6\,\mu$m coverage.  This
candidate is located near a bright galaxy, where the photometry in the
IRAC band is challenging with aperture photometry (Fig.~\ref{fig:jh}).
{To extract the flux from the candidate, we use TPHOT
  \citep{merlin16}, which models the IRAC flux based on the light
  profile obtained from the high-resolution \hst\ images.  We use the
  \jh+\hh\ image as the reference model and a convolution kernel that
  is constructed from the PSFs in the \hh\ and IRAC $3.6\mu{\rm m}$
  bands.  This approach yields an extracted model magnitude of
  $m_{3.6}=23.8\pm0.7$ for the candidate.} By including this IRAC flux in
the photometric redshift code, the low-$z$ probability peak is
$\sim16\%$ of the high-$z$ peak, and thus we retain the galaxy as the
final, and the only, $z\sim10$ candidate in this study
(Fig.~\ref{fig:jh}).

One of the $z\sim9$ candidates, $0956+2848$-98, has IRAC $3.6\,\mu$m
coverage, but shows no detection ($S/N<1$), and a $1\,\sigma$ lower
limit is available from the rms map ($>25.5$\,mag;
Fig.~\ref{fig:yjh}).  The photometric redshift of the source including
the IRAC upper limit remains unchanged, and thus we retain this in the
sample of final $z\sim9$ candidates.

\subsection{Additional F098M photometry}\label{ssec:f098m} {One of
  our $z\sim9$ candidates is by chance overlapping with previous
  archived observations, both from the Hubble Infrared Pure Parallel
  Imaging Extragalactic Survey (HIPPIES) pure-parallel program (GO
  11702; PI: Yan; $\sim10$\,orbits) and from a previous BoRG campaign
  ($\sim1$\,orbit).  Combining all data, and in particular thanks to
  the very deep GO 11702 exposure, the object shows a significant
  detection in the deep F098M image, $m_{098}=27.1\pm0.2$.  The most
  likely photometric redshift for this object becomes $z\sim2.2$, and
  thus we exclude it from the final candidate sample.}  {Two other
  $z\sim9$ candidates, which were reported in C16, have also been
  recently followed up with the same filter
  \citep[1\,orbit;][]{livermore18}.  The follow-up data revised the
  redshift to $z\sim1.8$ for one of the two galaxies (while confirming
  the high-$z$ nature of the other), and thus this source is not included in our
  final sample.}

{This suggests that, even after the stringent selection of this
  study, low-$z$ interlopers could be selected as final high-$z$
  candidates. While our final candidates show low probability peaks at
  low $z$ ($\simlt16\%$), additional imaging around the putative Lyman
  break would sample the SED more accurately and further improve the
  quality of the high-$z$ photometric candidates in the absence of
  spectroscopic confirmation. Deep \spit\ data would constrain the
  rest-frame optical wavelength range, but only if sources are well
  isolated (see the case for $2140+0241$-303). Space-based deep
  medium-/narrowband observations, as in the \borg\ Cycle 25 program
  \citep{livermore18}, are ideal before the advent of next-generation
  facilities such as \jwst\ (see also Appendix~\ref{sec:Ac}).  }

\subsection{Absolute Magnitude}\label{ssec:ab}
The UV absolute magnitude at rest-frame $1450$\,\AA, $M_{UV}$, is then
calculated based on the photometric redshift (i.e. the $k$-correction) and scaled {\hh\ magnitude
(Section~\ref{ssec:cs}).} We assume log-normal SED with a UV slope
$\beta=-2.0$ ($f_{\lambda}\propto\lambda^{\beta}$), appropriate for the
high-$z$ population \citep[e.g.,][]{fan03, bouwens14}.
We checked that our choice of $\beta$ from observed values \citep[$\sim-1.5$ to $-2.0$ for
$M_{UV}=-23$ to $-20$;][]{bouwens14} does not change our final result.

\subsection{Magnification bias}\label{ssec:mag}

Among the sample, three candidates have neighboring foreground objects in each FoV.
The observed light from those high-$z$ candidates is thus possibly affected by the gravitational potential of the foreground objects.

We calculate the magnification by foreground objects in the same manner as \citet{mason15}.
Briefly, the photometric redshift of low-$z$ objects is derived with \eazy.
The probability distributions for both
foreground and high-$z$ candidates, and their separation, are then used to estimate the
magnification. 
Single isothermal spheres are assumed for the mass profile of foreground deflectors.
The Einstein radius of deflectors is estimated from photometry using a redshift-dependent relation of \citet{faber76}.
The resulting magnifications are $1.5_{-0.3}^{+0.7}$ (2140+0241-303), $1.1_{-0.1}^{+0.1}$ (0751+2917-499), and $1.7_{-0.4}^{+0.5}$ (2229-0945-394).
The error in magnification is also integrated into the calculation of absolute magnitude.

\begin{deluxetable*}{lcrccccccccccccc}
\tabletypesize{\footnotesize}
\tablecolumns{16}
\tablewidth{0pt} 
\tablecaption{Coordinates and photometric properties of final $z\sim10$ and $z\sim9$ candidates}
\tablehead{\colhead{Object ID} & \colhead{$\alpha$(J2000)} & \colhead{$\delta$(J2000)} & \colhead{\hh} & \multicolumn{2}{c}{Colors} & \multicolumn{6}{c}{S/N} & \multicolumn{2}{c}{$z_{\rm phot.}$} & \colhead{$M_{\rm UV}$$^{\dagger}$} & \colhead{$r_{50}$$^{\dagger}$}\\
\colhead{} & \colhead{(deg)} & \colhead{(deg)}& \colhead{(mag)} &\colhead{$C_1$} & \colhead{$C_2$}& \colhead{$O_{\rm ap.}$}& \colhead{\oo} & \colhead{\yy} & \colhead{\jj} & \colhead{\jh} & \colhead{\hh} & \colhead{flat} & \colhead{prior} & \colhead{(mag)} & \colhead{(arcsec)}
}
\startdata
\cutinhead{$z\sim10$ candidates}
2140+0241-303 & 324.885438 & 2.685170 & $24.4$ & $1.7\pm0.3$ & $0.5\pm0.1$ & $0.3$ & $0.5$ & $0.9$ & $3.1$ & $10.2$ & $15.8$ & 10.0 & 10.0 & $-22.6_{-0.2}^{+0.4}$ & 0.18 \\

\cutinhead{$z\sim9$ candidates}
0956+2848-98 & 149.122734 & 28.792008 & $26.6$ & $1.7\pm0.8$ & $-0.2\pm0.2$ & $-0.2$ & $-0.2$ & $1.4$ & $5.7$ & $7.6$ & $5.8$ & 8.8 & 8.8 & $-20.7_{-0.2}^{+0.1}$ & 0.17 \\
2229-0945-394 & 337.190253 & -9.749134 & $25.0$ & $>2.6$ & $0.2\pm0.2$ & $0.7$ & $0.1$ & $0.4$ & $7.4$ & $8.7$ & $9.3$ & 9.0 & 9.0 & $-21.8_{-0.3}^{+0.3}$ & 0.16 \\

\enddata
\tablecomments{
$C_1$: \jj-\hh\ and \yy-\jh\ colors for $z\sim10$ and $z\sim9$ candidates, respectively. 
$C_2$: \jh-\hh\ color for $z\sim10$ and $z\sim9$ candidates.
$O_{\rm ap.}$: Signal-to-noise ratio in aperture photometry (radius of 0.\!\arcsec16) of \oo\ band.
Column 8-12: Isophotal signal-to-noise ratio of each band. 
Column 13: Photometric redshift derived with a flat prior. Peak redshift is presented.
Column 14: Photometric redshift derived with the modeled prior. Peak redshift is presented.
Column 15: UV absolute magnitude at $z_{\rm mp}$ assuming $f_\lambda \propto \lambda^{-0.2}$.
Column 16: \sext\ half-light radius measured in \hh.
\\
$^{\dagger}$ Absolute magnitude and radius are corrected for the lens magnification (Section~\ref{ssec:mag}).
}
\label{tab:jh}
\end{deluxetable*}

\section{Final candidates}\label{sec:sample}
Through the selection processes described in Sections~\ref{ssec:cs} to \ref{ssec:spt}, we have collected one $z\sim10$ and two $z\sim9$ candidates (Figs.~\ref{fig:jh} and \ref{fig:yjh}).
{These candidates show extended morphology in \hh\ ($>0.\!\arcsec16$; compare $r_{\rm PSF}\sim0.\!\arcsec1$ for WFC3IR). 
In addition, their red colors in \yy$-$\jh\ assure that these are unlikely to be Galactic dwarf stars \citep{oesch13}.}
The photometric properties and redshifts of the final candidates are summarized in Table~\ref{tab:jh}, while those initially selected with the color selection but later rejected are presented in Appendix~\ref{sec:Ab}.
Here we look at the final candidates in detail before proceeding to estimation of the number density.

\subsection{Comparison to \citet{calvi16}}\label{ssec:comp}
Part of our data (36\,fields; $\sim40\%$) has been studied in C16.
While the initial color selections are identical in both studies, the application of our selection processes, in addition to the update in drizzle pipeline parameters and \sext\ parameters, may affect the final sample.
In what follows, we compare our candidates with those presented in C16.

\subsubsection{$z\sim10$ candidates}\label{sssec:jj}
Both C16's $z\sim10$ candidates are rejected in the present study
(as 2134-0708-645 and 2140+0241-38).  These objects have consistent
photometric properties between the two studies, including photometric redshift, though C16
used the BPZ photo-$z$ code \citep{benitez00, coe06}.  However, the first object has an inconsistent photo-$z$ after
the application of our phenomenological prior (2134-0708-645 in
Table~\ref{tab:rej}).  In fact, the probability distribution of this
object had a secondary peak at $z\sim4$ (see Figure~4 of C16;
$\sim40\%$ of the high-$z$ peak), and the rejection after the
application of our new prior is not surprising.
The second object has large apparent size ($r_{50}\sim0.\!\arcsec5$) and is excluded by the size criterion. 

\begin{figure*}
\centering
\input{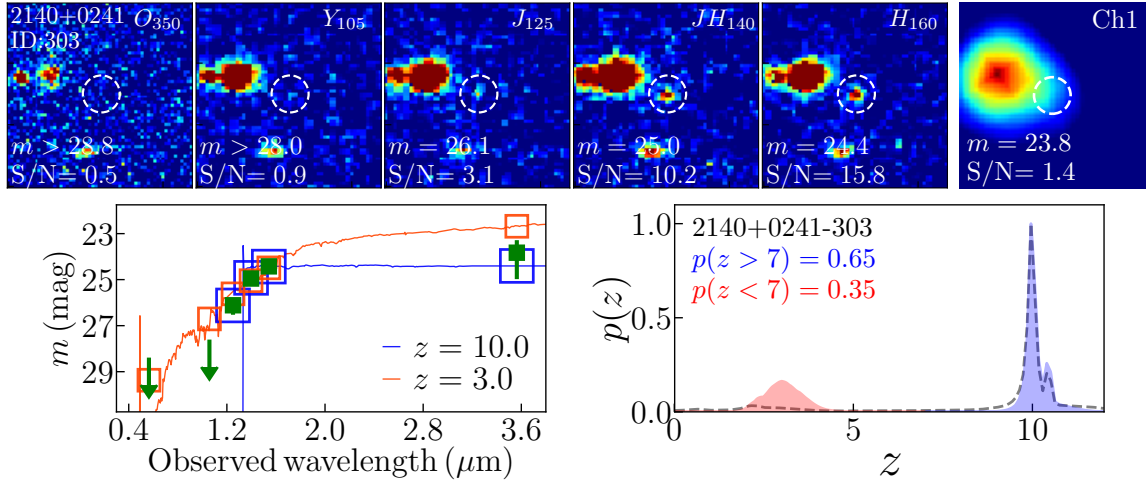}
\caption{
Final $z\sim10$ candidate (\jj\hh-dropout) that passes the color-cut criteria, visual inspection, and photometric redshift selection. 
{\it Top:} $30\times30$\,pixel ($\sim2.4$\,arcsec) postage stamps in five \hst\ filters and IRAC ch1.
The total magnitude and isophotal signal-to-noise ratio (S/N) are shown in each panel. 
$1\,\sigma$-limiting magnitudes are shown for those without detection (S/N$<1$).
{\it Bottom left:} spectral energy distribution.
Detection (S/N\,$>1$; green squares) and $1.5\,\sigma$ upper limit (arrows) are shown.
The best fit templates of $\chi^2$-minimization at high and low redshifts (blue and red solid lines, respectively) are shown, with expected fluxes at each observed band (open squares).
{\it Bottom right:} photometric redshift probability distributions (gray dashed line for likelihood; solid line for posterior). Total probabilities at $z>7$ and $z<7$ are also shown.
}
\label{fig:jh}
\end{figure*}

On the other hand, we find a new candidate (2140+0241-303; Fig.~\ref{fig:jh}), which was oddly not presented in C16.
We find that this object was rejected because of an infinity value in its flux measurement, which came from the artificial values ({\it NAN}) in the original rms map in C16. 
As described in Section~\ref{ssec:reduction}, we mask pixels with the artificial value when calculating their source fluxes for all images in this study. The treatment establishes this source as a viable $z\sim10$ candidate.

This candidate is much brighter ($H_{160}\sim24.4$\,mag) than any other candidates reported at $z>8$ so far \citep{coe13, zitrin15, oesch16, stefanon17, salmon18}.
Even after correction for magnification (Section~\ref{ssec:mag}), this candidate has an apparent magnitude of $\sim25$\,mag, corresponding to $M_{UV}\sim-22.7$\,mag, making it as one of the brightest objects at high $z$ \citep[see also Fig.3 of][]{salmon18}.
Without additional data, we cannot reject any contribution from non-stellar components, such as AGNs.

This candidate shows a small central blob in \yy, despite its non-detection with isophotal flux ($S/N = 0.9$).
As for $O_{350}$ (Section~\ref{ssec:cs}), we measure the $S/N$ in a small aperture ($r=0.\!\arcsec16$) in \yy, still finding no significant detection ($S/N\sim1.4$).
Since the exposure in this field is relatively shallow (two orbits), follow-up observations at $\sim1.1\mu{\rm m}$ would be extremely useful to elucidate this source. 

Another caveat is its clustering properties. By nature, bright objects are preferentially found in dense environments and thus neighboring objects are also expected
\citep{munoz08, trenti08, ishigaki16, harikane18}. However, with moderate depth of imaging, clustering is not expected to play a significant role in discriminating the nature of the candidate \citep[][]{ren18}. Not finding any dropout companions around this object is not unusual with our current imaging data. 
To further confirm any dropout clustering, deeper ($\Delta m\sim2$\,mag) follow-up observations would be required.

\begin{figure*}
\centering
\input{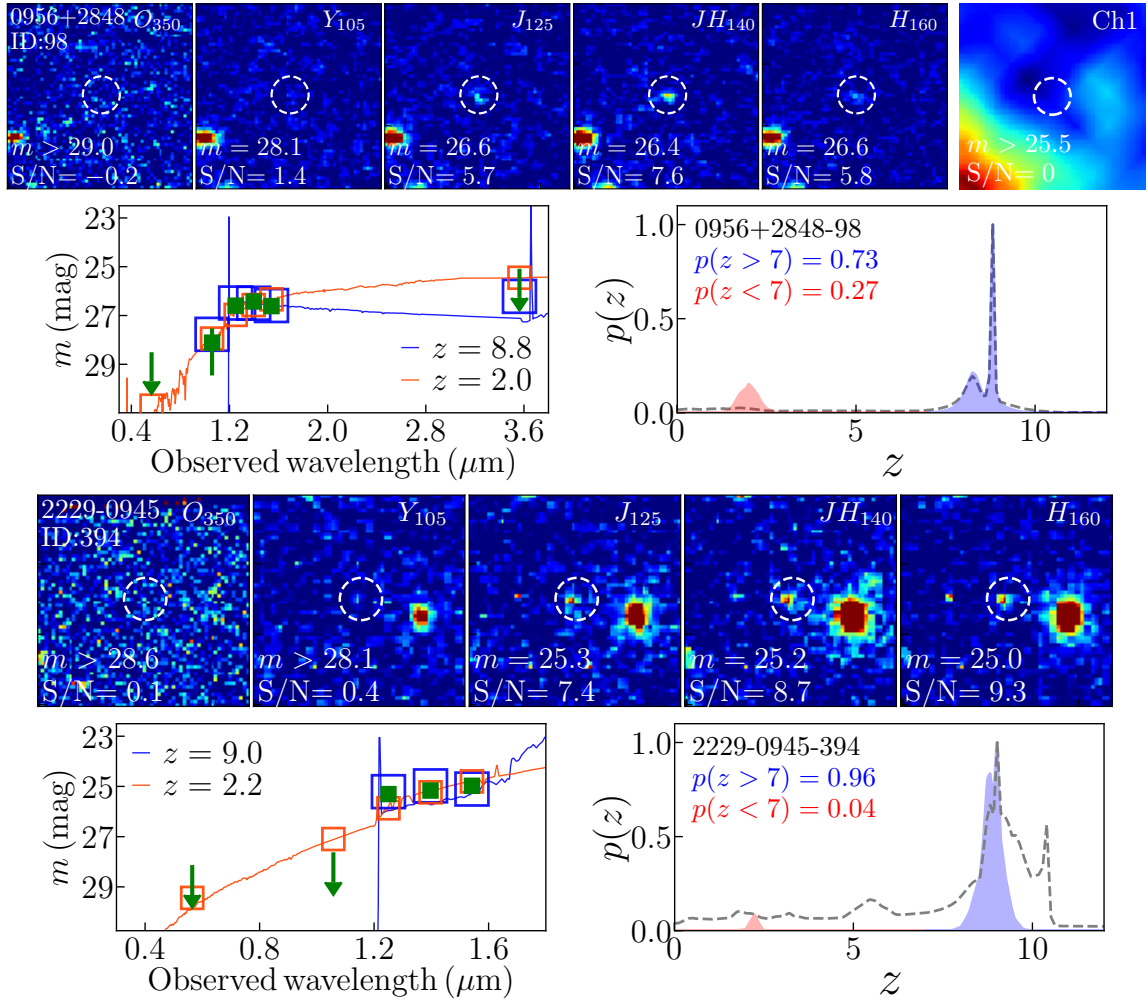}
\caption{
Same as Figure~\ref{fig:jh} but for final $z\sim9$ candidates (\yy\jh-dropouts).
}
\label{fig:yjh}
\end{figure*}


\subsubsection{$z\sim9$ candidates}\label{sssec:yy}
Two of three $z\sim9$ candidates in C16 are selected here (0956+2848-98 and 2229-0945-394) with consistent photometric properties, and successfully pass all the selection processes in the present study.

The third C16's candidate also satisfies all selection criteria in this study (0116+1425-442). 
However, a follow-up study with \hst\ F098M imaging
finds this object to be a low-$z$ interloper at $z\sim1.8$ \citep{livermore18}, as we described in Section~\ref{ssec:f098m}.

\section{Results}

\subsection{Effective Volume}
We calculate the effective volume by following \citet{oesch12} and C16
\citep[see also][for a public code based on the same approach adopted
here]{carrasco18}.  Briefly, we added artificial sources in empty
regions of each science frame.  Those sources were modeled with
realistic intrinsic distributions of UV colors and half-light radii,
and they had assigned redshifts and \hh-magnitudes to compute the
selection efficiency as a function of those parameters.

The sources were modeled with a mix of $n=1$ and $n=4$ \citet{sersic68} profiles, and with half-light radii as found in previous studies \citep{oesch10, grazian12, ono13, holwerda15, kawamata15, curtis-Lake16, bouwens17}.
We assumed that there is no significant evolution at $z\simgt8$ \citep[][]{wilkins16}, and modeled the SED with observed UV-slope $\beta$ fixed to the $z\sim8$ value \citep[][]{bouwens14}.
We calculated the completeness as a function of apparent \hh\ magnitude, $C(m)$, and the source selection function as a function of magnitude and redshift, $S(z,m)$, in each sightline by detecting those artificial sources.

The effective comoving volume is then calculated as
\begin{equation}
V_{\rm eff}(m) = \int_{0}^{\infty} S(z,m) C(m) {dV\over{dz}} dz,
\end{equation}
where $dV/dz$ represents differential comoving volume at redshift $z$.
The total volume, the sum of the effective volume over all the effective survey fields, i.e. the region that is not occupied by bright sources, ranges from $\sim4\times10^{4}$\,Mpc$^{3}$ to $\sim9\times10^{5}$\,Mpc$^{3}$ for our magnitude range, $-24\simlt M_{UV}\simlt-20$.
The effective volume used for each magnitude bin is summarized in Table~\ref{tab:LF}.

\subsection{Number density of high-$z$ candidates}\label{ssec:nd}

With the final candidates and effective volume derived in the previous sections, we estimate the stepwise number density.
The density is estimated by dividing the number of candidates at a given UV absolute magnitude by the effective volume. 
The 1\,$\sigma$ confidence level is estimated by assuming a Poisson distribution \citep[][]{gehrels86}.
For magnitude bins with no candidates, we show $1\,\sigma$ upper limits derived from the Poisson distribution and effective volume.

We also take account of contamination by multiplying by $(1-f_{\rm cont})$, where $f_{\rm cont}=0.2$ is the contamination fraction defined in Section~\ref{ssec:pz} \citep[see also][]{bradley12, schmidt14b}.
With this, the uncertainty from this arbitrary contamination fraction is now limited to $\simlt 0.1$\,dex in the estimation of the number density.

\begin{figure*}
\centering
\includegraphics[width=0.9\textwidth]{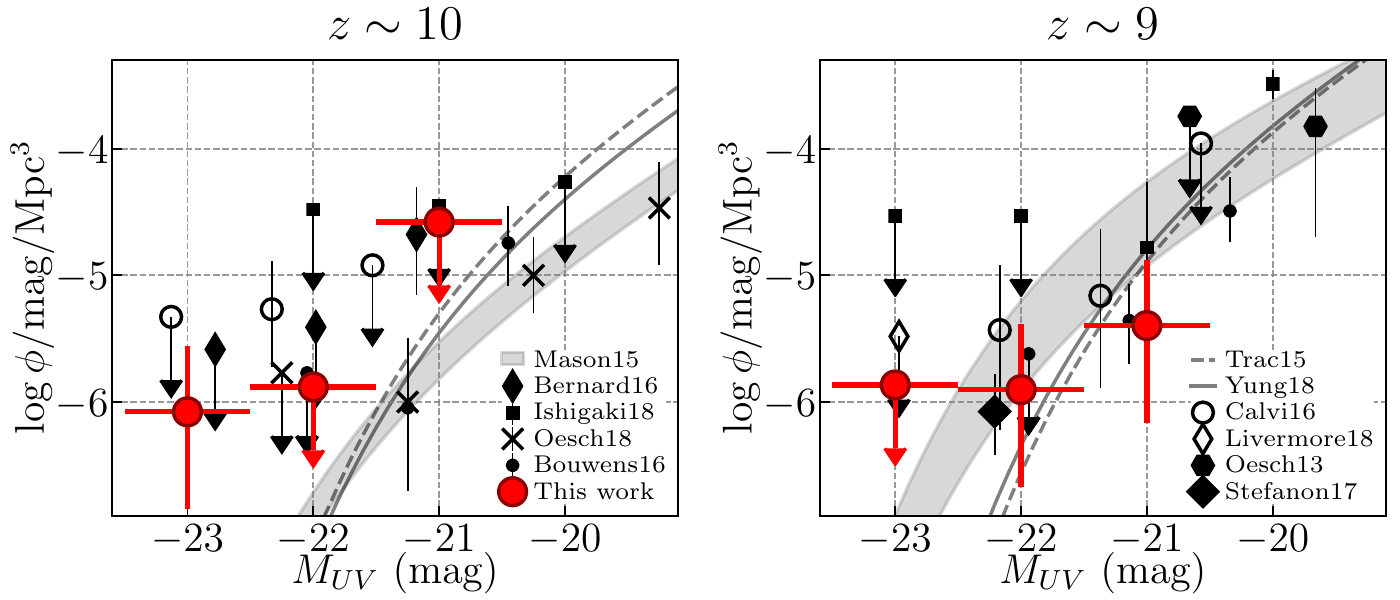}
\caption{
Number density of $z\sim10$ (left) and $z\sim9$ (right) candidates derived in this study.
The bin size in magnitude ($\Delta M=0.5$) is taken so that it becomes larger than the typical errors from photometric redshift and magnitude.
Previous results from the partial \borgz\ data are shown with open symbols (\citealt[][circles]{calvi16}; \citealt[][diamonds]{livermore18}).
Other results from previous observations in this magnitude range are also shown with filled symbols (\citealt[][hexagons]{oesch13}; \citealt[][diamonds]{bernard16}; \citealt[][large diamond]{stefanon17}; \citealt[][squares]{ishigaki18}; \citealt[][dots]{bouwens15}; \citealt[][crosses]{oesch18}).
Theoretical expectations of luminosity functions at $z\sim10$ and 9 from \citet[][gray shaded regions]{mason15b}, \citet[][dash lines]{trac15}, and \citet[][solid lines]{yung18} are shown.
}
\label{fig:LF}
\end{figure*}

Our results are summarized in Table~\ref{tab:LF} and shown in Figure~\ref{fig:LF}, where we find consistency with previous studies in \borg\ (C16) and other fields \citep{oesch13, bouwens15, oesch18}.
Our upper limits constrain the number density at values a factor of $\sim0.3$\,dex lower than in C16, thanks to the volume from 79\,independent fields.
It is noted that since three of C16's candidates are rejected as low-$z$ interlopers in this study, one of C16's data points at $z\sim10$ ($M_{UV}\sim-22.3$) now becomes an upper limit.

At $z\sim9$, we find a good agreement with theoretical expectations at $M_{UV}\simgt-22.5$, both semi-analytical models \citep[e.g.,][]{mason15b} and cosmological simulation \citep[e.g.,][]{trac15, cowley18, yung18}.

At the brightest magnitude bin of $z\sim10$, however, the estimated
density is $\simgt1$\,dex above what theoretical models expect.  While
our sample size is limited, and follow-up studies are necessary to
differentiate those candidates from the contamination, this may
highlight factors that were missed in previous models, such as
evolution of the shape of LFs.  The contribution of AGNs, which
possibly boost the observed light of our candidates, would also be
worth investigating.

\section{Discussion}\label{sec:dis}

\subsection{Impact on previous LFs at $z\sim9$}\label{ssec:lf}

Our magnitude range is limited to the bright end, and thus deriving
robust LFs with the present data alone is challenging.  Still, it is
worth investigating how much our new candidates influence LFs derived
in previous studies that focused on faint objects but with limited volume,
where bright objects could be easily missed.  We attempt here to
combine our $z\sim9$ candidates with those in \citet{ishigaki18} and
investigate the impact on the best-fit parameters.\footnote{We adopt the result of
  \citet{ishigaki18} because it consists of the largest survey
  volume down to $M_{UV}\sim-13$\,mag at $z\sim9$. It should therefore be
  least affected by adding our bright-end data points,
  which is ideal for our test here.}

The $z\sim9$ candidates in \citet{ishigaki18} are selected by color-color selections from all the {\it Hubble Frontier Fields} ($\sim56$\,arcmin$^2$ in total), similar to our scheme. We take the data points of stepwise LFs in their Figure~4, which already takes into account the effect of lens magnification.
While their samples include those with photometric redshifts inconsistent with the selection, we still use their values to reproduce their best-fit parameters of LF.
The error, dominated by Poisson error, is recalculated in the same manner as ours.

We first fit only their points with a similar technique, Markov Chain
Monte Carlo (MCMC) method, to see if our fitting method reproduces
their Schechter function fit:
\begin{equation}
\begin{aligned}
\phi(M_{UV}) = {{\ln 10} \over{2.5}} \phi^* \times 10^{0.4(\alpha+1)(M_{UV}-M_{UV}^*)}\\
 \times {\rm exp}\big[-10^{-0.4 (M_{UV}-M_{UV}^*)}\big]
\end{aligned}
\end{equation}

We use the {\tt{emcee}} code \citep{foreman13}, assuming flat priors in the following ranges: $\alpha\in[-10:0]$, $M^*\in[-30:0]$/mag, and $\log \phi^* \in[-10:0]$/Mpc$^{-3}$. The calculation is repeated for $N_{\rm mc1}=10000$ times, and the first half realization is discarded to avoid the bias from initial parameter values. The best-fit values and uncertainties (16/50/84th percentiles) are calculated from the rest of the chain.

When all the parameters are set as free, the fitting results are unconstrained with large uncertainty, as also seen in \citet{ishigaki18}, because the knee of the LF ($M_{UV}\simlt21$\,mag) is not sufficiently sampled by their data only. We then follow \citet{ishigaki18} and fix the two of the parameters ($\alpha=-1.96$ and $M_{UV}^*=-20.35$), finding consistent values for the others. This ensures that results with our fitting method are comparable.

We then add our data and fit with the Schechter function, but with a few updates in the treatment of uncertainties.
We run the MCMC fitting routine introduced above for $N_{\rm mc2}=300$ times by fluctuating the data, as described in the following.

First, we fluctuate the observed flux within the random flux error and the systematic uncertainty from the magnification model in each iteration.
While the latter uncertainty is relatively small for our candidates, some of those of \citet{ishigaki18} have large uncertainties in the magnification (by a factor of $\sim3$) because of complicated cluster lens modeling.\footnote{For the candidates of \citet{ishigaki18}, we use uncertainties in photometric redshift and magnification listed in \citet{kawamata18}.}

Second, we take account of the redshift uncertainty. 
In \citet{ishigaki18}, they did not have a selection criterion with photometric redshifts, while our final sample does include a photometric redshift selection (Section~\ref{ssec:pz}). To make the two samples consistent, we fluctuate redshifts within their $1\,\sigma$ uncertainty range in each iteration and select those with $7.7<z<9.7$ ($95\%$ confidence interval for our $z\sim9$ candidates) when calculating the number density.
Absolute magnitudes are also recalculated based on fluctuated redshifts in this step.

Finally, we refine the UV magnitude grid when calculating the number density, because different grids change the stepwise number density and can affect the fitting parameters \citep[][]{schmidt14b}.
We set the magnitude grid as $M_{UV}\in[-24+rand(-0.5,0.5):-13+rand(-0.5,0.5)]$ with a magnitude bin size $\Delta M_{UV}=rand(0.3,1.5)$ in each iteration, where $rand(a,b)$ is a random float value taken from the range between $a$ and $b$.
The effective volume is interpolated to the refined magnitude grid.
Skipping the third process would underestimate the uncertainty down to $\sim30\%$, depending on the size of the magnitude bin.

\begin{figure*}
\centering
\includegraphics[width=0.45\textwidth]{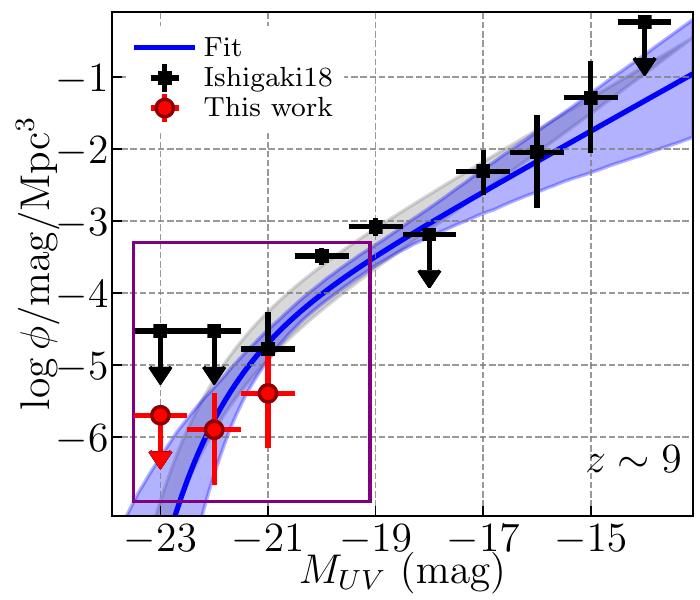}
\includegraphics[width=0.4\textwidth]{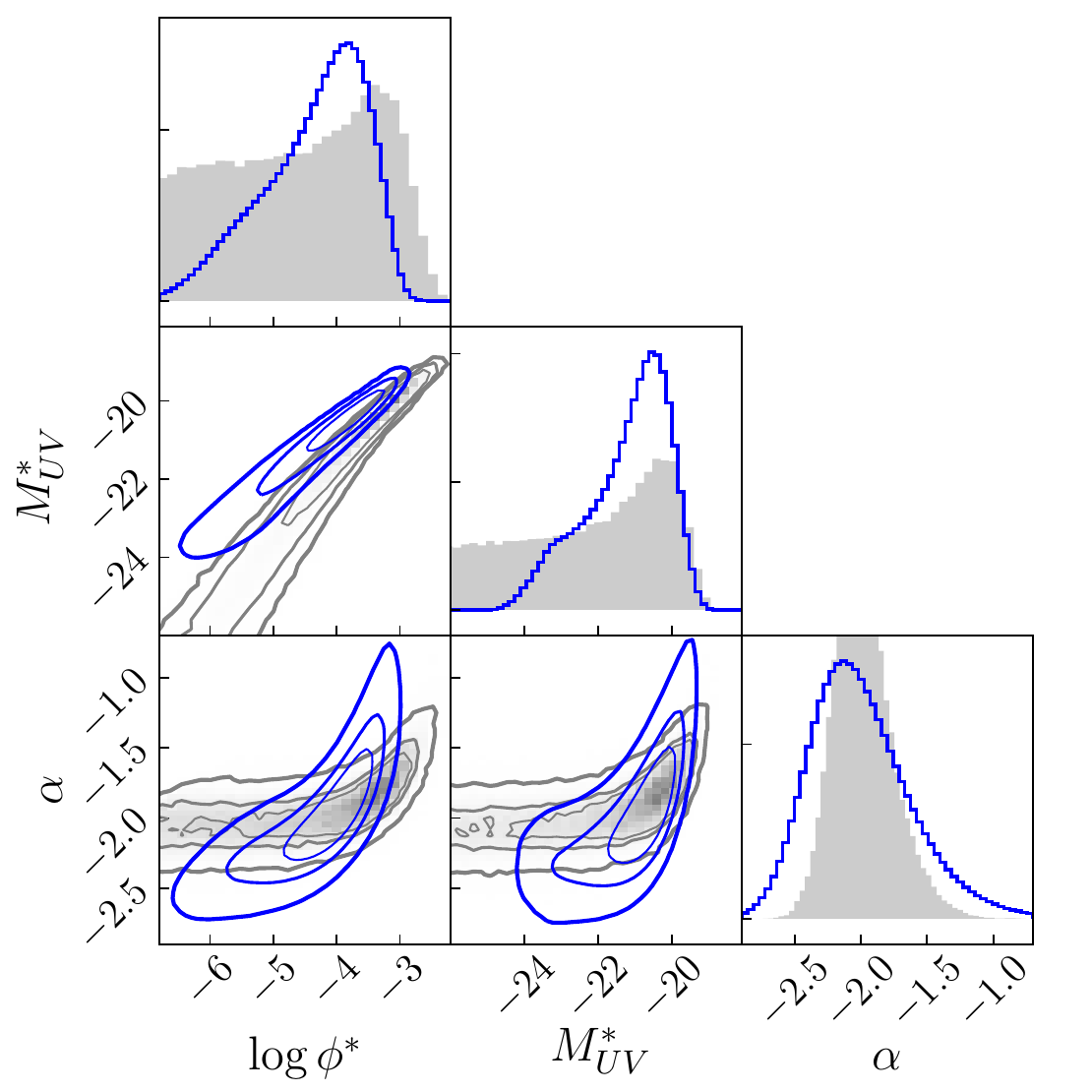}
\caption{
Left: LF fit with the Schechter function form to the combined data from this study (red circles) and \citet[][black squares]{ishigaki18}.
The fit with the best-fit parameters is shown with blue solid line, with its 16/84th percentiles from all the realizations (blue filled region).
The theoretical LF by \citet{mason15b} is shown for comparison (gray shaded region).
The zoomed region shown in Figure~\ref{fig:LF} is framed with purple lines.
Right: Covariance matrices of the LF parameters.
Fitting only with previous data from \citet{ishigaki18} shows unconstrained results in parameter spaces (filled gray histograms and contours; 39/68/95\%), whereas inclusion of our data shows constraints for all parameters (blue contours and histograms).
}
\label{fig:LFcomp}
\end{figure*}

The result from combining our new observations with the literature
data is shown in Fig.~\ref{fig:LFcomp}, where we find that all three
parameters are constrained within physically meaningful ranges
(Table~\ref{tab:LFfit}).  We present the 50/16/84th percentiles taken
from the synthesized MCMC chain ($N_{\rm mc1}\times N_{\rm mc2}$) as
the best-fit parameters.  Compared to the fixed values of \citet{ishigaki18} ($\alpha=-1.96$ and $M_{UV}^*=-20.35$), our value for $\alpha$
is smaller ($-2.1$), and $M_{UV}^*$ is $\sim0.6$\,mag smaller
($-21.0$), which can be understood from the fact that our data constrain
the bright end at a lower number density than previously possibly.

With the best-fit Schechter parameters, we estimate the cosmic UV luminosity density, 
\begin{equation}
\rho_{UV} = \int_{-\infty}^{M_{\rm lim.}} dM_{UV} L(M_{UV}) \phi(M_{UV})
\end{equation}
where $L(M_{UV})$ is the UV luminosity at a given UV absolute magnitude.
As summarized in Table~\ref{tab:LFfit}, we find $\log\rho_{UV}\sim25.3$ ergs$^{-1}$Hz$^{-1}$Mpc$^{-3}$ ($25.6$) with $M_{\rm lim}=-17$ ($-15$), or $\log \psi_{*}\sim-2.6$\,$M_\odot$yr$^{-1}$Mpc$^{-3}$ ($-2.4$) when the conversion in \citet{madau14} is applied. 

The derived value is consistent with that derived in \citet[][$\log \psi_{*}\sim-2.7$\,$M_\odot$yr$^{-1}$ with $\alpha=-1.96$ and $M_{UV}^*=-20.35$]{ishigaki18}, or slightly larger than that in \citet[][$\log\psi_*\sim-2.9$ with $\alpha=-1.73$ and $\log \phi^*=-2.94$]{oesch13}, but in excellent agreement with \citet{mcleod16} and theoretical models \citep[e.g.,][see also Figure 5 of \citealt{oesch18}]{tacchella13, mason15}.
The luminosity density derived here lies between those claimed at lower redshifts \citep[e.g., $z\sim8$ in][]{bouwens15} and higher redshift \citep[$z\sim10$ in][]{oesch18}.

Our updated value of the UV luminosity density does not qualitatively
change the conclusion in \citet{ishigaki18}, where they found a linear
relation in $z$-$\log \rho_{UV}$ up to $z\sim9$ \citep[see
also][]{oesch13,bouwens16,mcleod16}, which may then turn into an
accelerated decrease at yet higher redshift \citep[][]{oesch18}.

We also fit the number density with a double power law, motivated by recent studies at lower redshift \citep[e.g.,][]{ono18}:
\begin{align*}
\phi(M_{UV}) = {{\ln 10} \over{2.5}} \phi^* \times \big[10^{0.4(\alpha+1)(M_{UV}-M_{UV}^*)}\\
 + 10^{0.4(\beta+1)(M_{UV}-M_{UV}^*)}\big]^{-1}
\end{align*}

We set the same prior range as above for $\alpha$, $M_{UV}^*$, and $\phi^*$, and set $\beta\in[-10:\alpha]$.
The fit shows unconstrained $\beta$ and large uncertainties for the other parameters when all the parameters are free.
When the characteristic magnitude is fixed ($M^*=-20.35$), the other parameters return some constrained values, but still with relatively large uncertainties ($\log \phi^*=-3.48_{-0.11}^{+0.11}$, $\alpha=-1.92_{-0.17}^{+0.17}$, and $\beta=-5.80_{-1.89}^{+2.00}$).
Thus, we conclude that we are still unable to rule out this functional
form, and that additional large-area surveys would be highly beneficial.

\begin{deluxetable}{cccc}
\tabletypesize{\footnotesize}
\tablecolumns{4}
\tablewidth{0pt} 
\tablecaption{Number density of dropout candidates}
\tablehead{
\colhead{$M_{UV}$} & \colhead{$\log\phi$} & \colhead{Number} & \colhead{$V_{\rm eff}$}\\
\colhead{(mag)} & \colhead{(1/Mpc$^{3}$/mag)} & \colhead{} & \colhead{($10^{4}$\,Mpc$^{3}$)}
}
\startdata
\cutinhead{$z\sim10$}
$-23.0$ & $-6.1_{-0.8}^{+0.5}$ & $1$ & $95.68$\\
$-22.0$ & $<-5.9$ & $0$ & $76.08$\\
$-21.0$ & $<-4.6$ & $0$ & $3.80$\\

\cutinhead{$z\sim9$}
$-23.0$ & $<-5.9$ & $0$ & $73.39$\\
$-22.0$ & $-5.9_{-0.8}^{+0.5}$ & $1$ & $63.87$\\
$-21.0$ & $-5.4_{-0.8}^{+0.5}$ & $1$ & $19.94$\\

\enddata
\tablecomments{
Errors of number densities are dominated by Poisson uncertainty.
}
\label{tab:LF}
\end{deluxetable}

\begin{deluxetable*}{ccccccc}
\tabletypesize{\footnotesize}
\tablecolumns{7}
\tablewidth{0pt} 
\tablecaption{Fitting Parameters for Schechter Luminosity Function and Cosmic Star Formation Rate Density at $z\sim9$}
\tablehead{
\colhead{$\log \phi^*$} & \colhead{$M_{UV}^*$} & \colhead{$\alpha$} & \multicolumn{2}{c}{$\log\rho_{\rm UV}$} & \multicolumn{2}{c}{$\log\psi_*$}\\
\colhead{(Mpc$^{-3}$\,mag$^{-1}$)} & \colhead{(mag)} & \colhead{} & \multicolumn{2}{c}{(erg\,s$^{-1}$\,Hz$^{-1}$\,Mpc$^{-3}$)} & \multicolumn{2}{c}{($M_\odot$\,yr$^{-1}$\,Mpc$^{-3}$)} \\
 &  &  & \colhead{($M_{UV}<-17$)} &  \colhead{($M_{UV}<-15$)} & \colhead{($M_{UV}<-17$)} &  \colhead{($M_{UV}<-15$)} 
}
\startdata
$-4.24_{-0.93}^{+0.56}\ _{-0.38}^{+0.44}$ & $-21.01_{-1.35}^{+0.69}\ _{-0.50}^{+0.58}$ & $-2.06_{-0.29}^{+0.31}\ _{-0.15}^{+0.27}$ & $25.33_{-0.12}^{+0.01}\ _{-0.09}^{+0.09}$ & $25.55_{-0.24}^{+0.10}\ _{-0.15}^{+0.15}$ & $-2.61_{-0.12}^{+0.01}\ _{-0.09}^{+0.09}$ & $-2.39_{-0.24}^{+0.10}\ _{-0.15}^{+0.15}$ \\

\enddata
\tablecomments{
50th and 16/84th percentiles are taken from the MCMC realization as the best-fit values and their uncertainties.
Associated errors are random photometric errors and systematic errors from the binning size in the UV absolute magnitude.
We use $\kappa_{UV}=1.15\times10^{-28}$\,$M_\odot$\,yr$^{-1}$/erg\,s$^{-1}$\,Hz$^{-1}$ \citep{madau14} to convert from the UV luminosity density ($\rho_{UV}$) to the star formation rate density ($\psi_*$).
}
\label{tab:LFfit}
\end{deluxetable*}


\section{Summary}
\label{sec:summary}
In this paper, we presented the full data set of \borgz, which
collected five-band imaging with \hst/WFC3IR and UVIS from 79
independent sightlines ($\sim370$\,arcmin$^2$), and we analyzed the data to investigate the number density of bright galaxies at $z\sim10$ and $9$.
Our findings are summarized as follows.
\begin{itemize}
\item With the standard color selection, we first found \njjcol\ and \nyycol\ candidates at $z\sim10$ and 9, respectively. Through the more strict selection process with photometric redshift and \spit\ photometry, we then identified respectively one and two sources as the final candidates (Figs.~\ref{fig:jh} and \ref{fig:yjh} and Table~\ref{tab:jh}).
\item The $z\sim10$ candidate is one of the brightest galaxy
  candidates at $z\simgt8$, an ideal target for follow-up observations.
\item We combined our data and previous faint candidates, and constrained the Schechter LF at $z\sim9$ without fixing any parameters for the first time.
\item From the best-fit LF parameters, we derived the UV luminosity density, $\log\rho_{UV}\sim25.3$\,ergs$^{-1}$Hz$^{-1}$Mpc$^{-3}$ (when integrated at $M_{UV}<-17$), which is consistent with an unaccelerated decreasing of the luminosity density from lower redshift.
\end{itemize}

Throughout the present study, it became clear that the selection of
high-$z$ candidates still suffers from low-$z$ interlopers, especially
with a standard color selection method (see also \citealt{vulcani17}).
Thanks to our survey design, these candidates are bright and ideal for
follow-up photometric and spectroscopic observations that would
require only a small fraction of the time devoted to the blind initial
search.  Future facilities with larger fields of view, such as WFIRST
\citep{spergel15}, will increase the number of candidates at a similar
redshift range and luminosity.  Its grism spectroscopic observations,
or observations at longer wavelength by \jwst\ \citep[e.g.,][]{kalirai18}, will be necessary to
improve the sample quality beyond that possible with a dropout
selection. In addition, such follow-up would also determine the physical
properties of the objects such as stellar masses, ages, and possibly
chemical composition.  Given the limited lifetime of \jwst, a
BoRG-like pure-parallel observing strategy will be an efficient way of
searching for the first bright galaxies at $z>8$ both with \hst\ and
with \jwst.

\acknowledgements 
We thank the anonymous referee for providing constructive comments.
We thank Masafumi Ishigaki
for providing the data points of luminosity function in
\citet{ishigaki18}. T.M. thank Yuichi Harikane for fruitful discussion.
Based on observations made with the NASA/ESA Hubble
Space Telescope, which is operated by the Association of Universities
for Research in Astronomy, Inc., under NASA contract NAS
5-26555. These observations are associated with programs 14701 and
13767. This research was conducted by the Australian Research Council
Centre of Excellence for All Sky Astrophysics in 3 Dimensions (ASTRO
3D), through project number CE170100013. 
C.M. acknowledges support provided by NASA through the NASA Hubble Fellowship grant HST-HF2-51413.001-A awarded by the Space Telescope Science Institute, which is operated by the Association of Universities for Research in Astronomy, Inc., for NASA, under contract NAS5-26555.

\clearpage
\appendix

\section{A. Survey fields and limiting magnitudes}\label{sec:Aa}
We list here exposure times and limiting magnitudes of all survey fields in the BoRG Cycle 22 observation.
The $5\,\sigma$ limiting magnitudes are calculated from rms maps. We measure the median rms value with an $0.\!\arcsec32$ radius aperture in empty sky regions.
Ten out of 89 original fields are discarded, either because the fields are excessively dominated by stars or the acquisition of a guide star failed.
Exposure times and limiting magnitudes for all fields are summarized in Table~\ref{tab:field}.

\begin{deluxetable*}{lrrccccccccccccc}
\tabletypesize{\footnotesize}
\tablecolumns{14}
\tablewidth{0pt} 
\tablecaption{BoRG fields coordinates, exposure times, and $5\,\sigma$ limiting magnitudes}
\tablehead{\colhead{Field ID} & \colhead{$\alpha$(J2000)} & \colhead{$\delta$(J2000)} & \colhead{No. of}  & \colhead{E(B-V)}  & \multicolumn{2}{c}{F350LP} & \multicolumn{2}{c}{F105W} &\multicolumn{2}{c}{F125W} & \multicolumn{2}{c}{F140W} & \multicolumn{2}{c}{F160W} & \colhead{$A_{\rm e}$}\\
        \colhead{}& \colhead{(deg)} & \colhead{(deg)}& \colhead{orbits} && \colhead{t$_{\rm exp}$}  & \colhead{m$_{\rm lim}$} & \colhead{t$_{\rm exp}$} & \colhead{m$_{\rm lim}$} & \colhead{t$_{\rm exp}$} & \colhead{m$_{\rm lim}$} & \colhead{t$_{\rm exp}$} & \colhead{m$_{\rm lim}$} & \colhead{t$_{\rm exp}$} & \colhead{m$_{\rm lim}$} & \colhead{($\square \arcmin$)}}
\startdata
0058-7201$^\ast$ & 1.458e+01 & -7.201e+01 & 2 & 0.322 & --- & --- & --- & --- & --- & --- & --- & --- & --- & --- & ---\\
0111-7248$^\ast$ & 1.767e+01 & -7.280e+01 & 4 & 0.087 & --- & --- & --- & --- & --- & --- & --- & --- & --- & --- & ---\\
0116+1425 & 1.906e+01 & 1.441e+01 & 3 & 0.041 & 2408 & 26.41 & 1758 & 26.41 & 2058 & 26.54 & 2208 & 26.43 & 2095 & 27.04& 4.61\\
0119-3411 & 1.968e+01 & -3.418e+01 & 3 & 0.027 & 1758 & 26.05 & 1305 & 26.11 & 1505 & 25.87 & 1605 & 26.08 & 1306 & 26.96& 4.62\\
0132+3035$^\ast$ & 2.311e+01 & 3.059e+01 & 3 & 0.049 & --- & --- & --- & --- & --- & --- & --- & --- & --- & --- & ---\\
0132-7326 & 2.305e+01 & -7.344e+01 & 2 & 0.073 & 1255 & 26.27 & 1455 & 26.68 & 1705 & 26.46 & 1805 & 26.30 & 1695 & 27.10& 4.64\\
0133+3043$^\ast$ & 2.337e+01 & 3.072e+01 & 3 & 0.041 & --- & --- & --- & --- & --- & --- & --- & --- & --- & --- & ---\\
0134+3034$^\ast$ & 2.348e+01 & 3.057e+01 & 3 & 0.043 & --- & --- & --- & --- & --- & --- & --- & --- & --- & --- & ---\\
0134+3041$^\ast$ & 2.343e+01 & 3.068e+01 & 2 & 0.041 & --- & --- & --- & --- & --- & --- & --- & --- & --- & --- & ---\\
0235-0357 & 3.880e+01 & -3.944e+00 & 16 & 0.023 & 11246 & 26.97 & 8235 & 27.23 & 9446 & 27.06 & 9846 & 27.14 & 8286 & 27.66& 6.35\\
0314-6712 & 4.843e+01 & -6.720e+01 & 8 & 0.038 & 7223 & 27.13 & 5129 & 27.23 & 6123 & 27.28 & 6223 & 27.40 & 4737 & 27.44& 5.07\\
0337-0507 & 5.437e+01 & -5.115e+00 & 3 & 0.044 & 2408 & 26.45 & 1708 & 26.61 & 2058 & 26.41 & 2108 & 26.38 & 1967 & 27.07& 4.76\\
0554-6005 & 8.839e+01 & -6.009e+01 & 4 & 0.057 & 2811 & 26.74 & 2058 & 26.97 & 2411 & 26.85 & 2511 & 26.44 & 2252 & 27.21& 4.62\\
0751+2917 & 1.177e+02 & 2.928e+01 & 4 & 0.043 & 2811 & 26.50 & 2008 & 26.68 & 2411 & 26.56 & 2461 & 26.53 & 2210 & 26.96& 4.62\\
0807+3606 & 1.219e+02 & 3.611e+01 & 4 & 0.049 & 3111 & 26.45 & 2308 & 26.67 & 2611 & 26.63 & 2711 & 26.23 & 2693 & 27.05& 4.64\\
0834+5238 & 1.285e+02 & 5.264e+01 & 3 & 0.035 & 1858 & 26.45 & 1355 & 26.64 & 1605 & 26.59 & 1705 & 26.69 & 1649 & 26.82& 4.64\\
0851+4240 & 1.327e+02 & 4.266e+01 & 3 & 0.025 & 2458 & 26.27 & 1808 & 26.38 & 2108 & 26.29 & 2208 & 26.27 & 2180 & 26.90& 4.61\\
0853+0310 & 1.332e+02 & 3.159e+00 & 3 & 0.050 & 1708 & 26.37 & 1255 & 26.60 & 1505 & 26.49 & 1555 & 26.61 & 1392 & 26.98& 4.61\\
0925+1360 & 1.413e+02 & 1.400e+01 & 3 & 0.032 & 1858 & 26.22 & 1305 & 26.42 & 1505 & 26.41 & 1705 & 26.50 & 1510 & 26.90& 4.62\\
0925+3439 & 1.413e+02 & 3.465e+01 & 3 & 0.020 & 2458 & 26.47 & 1758 & 26.52 & 2058 & 26.44 & 2158 & 26.54 & 2039 & 27.11& 4.64\\
0948+5757 & 1.470e+02 & 5.795e+01 & 3 & 0.015 & 2608 & 26.50 & 1908 & 26.70 & 2208 & 26.34 & 2308 & 26.50 & 2250 & 26.82& 4.61\\
0949+5759 & 1.473e+02 & 5.799e+01 & 4 & 0.014 & 3511 & 26.58 & 2561 & 26.84 & 2861 & 26.70 & 3111 & 26.78 & 2336 & 27.17& 4.64\\
0952+5150 & 1.480e+02 & 5.183e+01 & 3 & 0.007 & 1758 & 26.23 & 1958 & 26.49 & 2308 & 26.45 & 2358 & 26.52 & 1845 & 26.69& 4.62\\
0953+5150 & 1.483e+02 & 5.184e+01 & 4 & 0.009 & 2661 & 26.47 & 1958 & 26.67 & 2308 & 26.64 & 2358 & 26.68 & 1809 & 27.08& 4.64\\
0953+5153 & 1.483e+02 & 5.189e+01 & 8 & 0.009 & 5373 & 26.92 & 3917 & 27.14 & 4617 & 27.10 & 4717 & 27.16 & 3608 & 27.63& 4.95\\
0953+5157 & 1.483e+02 & 5.195e+01 & 4 & 0.010 & 2661 & 26.59 & 1958 & 26.90 & 2308 & 26.80 & 2358 & 27.00 & 1809 & 27.22& 4.61\\
0955+4528 & 1.488e+02 & 4.547e+01 & 3 & 0.011 & 1858 & 26.27 & 1355 & 26.45 & 1605 & 26.30 & 1705 & 26.26 & 1504 & 26.95& 4.64\\
0956+2848 & 1.491e+02 & 2.880e+01 & 6 & 0.018 & 4417 & 26.74 & 3214 & 26.85 & 3767 & 26.70 & 3864 & 26.69 & 2940 & 27.20& 4.76\\
1015+5945 & 1.537e+02 & 5.975e+01 & 6 & 0.010 & 4717 & 26.74 & 3467 & 26.87 & 4017 & 26.60 & 4214 & 25.67 & 3084 & 26.96& 4.69\\
1017-2052 & 1.544e+02 & -2.087e+01 & 3 & 0.044 & 1058 & 25.96 & 705 & 26.13 & 855 & 26.07 & 855 & 26.03 & 900 & 26.77& 4.64\\
1018+0544 & 1.545e+02 & 5.735e+00 & 3 & 0.020 & 2408 & 26.51 & 1758 & 26.70 & 2008 & 26.54 & 2108 & 26.56 & 2000 & 27.09& 4.64\\
1048+1518 & 1.620e+02 & 1.530e+01 & 3 & 0.028 & 2308 & 26.10 & 1658 & 26.16 & 1958 & 26.11 & 2058 & 26.24 & 1980 & 26.78& 4.61\\
1103+2913 & 1.657e+02 & 2.922e+01 & 4 & 0.029 & 3211 & 26.74 & 2311 & 26.81 & 2811 & 26.73 & 2911 & 26.77 & 2575 & 27.34& 4.64\\
1104+2813 & 1.660e+02 & 2.821e+01 & 4 & 0.033 & 2861 & 26.41 & 2058 & 26.50 & 2461 & 26.46 & 2508 & 26.50 & 2275 & 26.98& 4.63\\
1106+2925 & 1.665e+02 & 2.941e+01 & 3 & 0.030 & 2308 & 26.33 & 1605 & 26.43 & 1908 & 26.39 & 2008 & 26.43 & 1823 & 26.96& 4.66\\
1106+3508 & 1.665e+02 & 3.514e+01 & 4 & 0.018 & 3111 & 26.61 & 2208 & 26.79 & 2661 & 26.80 & 2761 & 26.61 & 2480 & 27.18& 4.64\\
1115+2548 & 1.687e+02 & 2.580e+01 & 4 & 0.017 & 2761 & 26.59 & 2008 & 26.81 & 2411 & 26.74 & 2461 & 26.80 & 2151 & 27.21& 4.62\\
1127+2653$^\dagger$ & 1.718e+02 & 2.688e+01 & 2 & 0.018 & --- & --- & --- & --- & --- & --- & --- & --- & --- & --- & ---\\
1136+0747 & 1.739e+02 & 7.786e+00 & 3 & 0.036 & 2408 & 26.43 & 1708 & 26.48 & 2058 & 26.57 & 2108 & 26.57 & 2113 & 27.06& 4.64\\
1142+2640 & 1.755e+02 & 2.667e+01 & 7 & 0.019 & 4370 & 26.71 & 3214 & 26.89 & 3717 & 26.87 & 3817 & 26.88 & 3166 & 27.21& 4.75\\
1142+2647 & 1.755e+02 & 2.678e+01 & 3 & 0.022 & 2158 & 26.37 & 1658 & 26.52 & 1908 & 26.33 & 1958 & 26.52 & 1486 & 26.90& 4.64\\
1142+3020 & 1.756e+02 & 3.034e+01 & 3 & 0.021 & 2408 & 26.52 & 1758 & 26.79 & 2108 & 26.71 & 2158 & 26.78 & 2130 & 27.20& 4.61\\
1143+3019 & 1.756e+02 & 3.032e+01 & 3 & 0.020 & 1808 & 26.35 & 1305 & 26.55 & 1505 & 26.36 & 1605 & 26.52 & 1266 & 26.88& 4.63\\
1149+2202 & 1.772e+02 & 2.203e+01 & 3 & 0.025 & 1758 & 26.19 & 1305 & 26.28 & 1555 & 26.30 & 1605 & 26.23 & 1480 & 26.74& 4.64\\
1152+3402 & 1.779e+02 & 3.403e+01 & 2 & 0.020 & 1605 & 26.38 & 1155 & 26.51 & 1405 & 26.44 & 1455 & 26.51 & 1154 & 26.91& 4.64\\
1152+5433 & 1.779e+02 & 5.455e+01 & 4 & 0.010 & 2811 & 26.43 & 2008 & 26.72 & 1758 & 26.52 & 2461 & 25.82 & 2185 & 27.04& 4.62\\
1152+5434 & 1.779e+02 & 5.456e+01 & 7 & 0.010 & 5070 & 26.64 & 3617 & 26.84 & 4320 & 26.82 & 4370 & 26.60 & 3947 & 27.33& 5.81\\
1154+4639 & 1.784e+02 & 4.665e+01 & 6 & 0.032 & 3717 & 26.81 & 2711 & 27.03 & 3211 & 26.91 & 3411 & 27.09 & 2583 & 27.46& 4.71\\
1160+0015 & 1.800e+02 & 2.522e-01 & 2 & 0.032 & 1805 & 26.51 & 1255 & 26.54 & 1505 & 26.44 & 1605 & 26.62 & 1473 & 26.88& 4.64\\
1209+4543 & 1.824e+02 & 4.572e+01 & 7 & 0.014 & 4420 & 26.94 & 3164 & 27.14 & 3717 & 27.12 & 3917 & 26.75 & 3500 & 27.48& 4.62\\
1218+3008 & 1.846e+02 & 3.013e+01 & 6 & 0.021 & 4067 & 26.73 & 2964 & 26.85 & 3417 & 26.80 & 3564 & 26.68 & 2764 & 27.20& 4.65\\
1229+0751$^\ast$ & 1.874e+02 & 7.857e+00 & 8 & 0.024 & --- & --- & --- & --- & --- & --- & --- & --- & --- & --- & ---\\
1313+1804$^\ast$ & 1.982e+02 & 1.807e+01 & 4 & 0.020 & --- & --- & --- & --- & --- & --- & --- & --- & --- & --- & ---\\
1334+3131 & 2.034e+02 & 3.152e+01 & 3 & 0.012 & 2408 & 26.52 & 1808 & 26.75 & 2108 & 26.68 & 2108 & 26.55 & 2133 & 27.15& 4.64\\
1410+2623 & 2.124e+02 & 2.638e+01 & 4 & 0.017 & 2811 & 26.63 & 2008 & 26.85 & 2411 & 26.85 & 2461 & 26.73 & 2210 & 27.21& 4.62\\
1413+0918 & 2.132e+02 & 9.299e+00 & 4 & 0.026 & 3061 & 26.65 & 2211 & 26.68 & 2561 & 26.71 & 2661 & 26.70 & 2358 & 27.12& 4.61\\
1421+4725 & 2.153e+02 & 4.741e+01 & 2 & 0.013 & 1205 & 26.21 & 1305 & 26.55 & 1605 & 26.38 & 1705 & 26.07 & 1480 & 26.93& 4.61\\
1431+0259 & 2.179e+02 & 2.989e+00 & 4 & 0.029 & 3011 & 26.48 & 2208 & 26.59 & 2561 & 26.52 & 2661 & 26.47 & 2551 & 27.08& 4.61\\
1437-0150 & 2.194e+02 & -1.826e+00 & 5 & 0.044 & 3364 & 26.57 & 2461 & 26.66 & 2914 & 26.63 & 3011 & 26.66 & 2541 & 27.15& 4.63\\
1438-0142 & 2.195e+02 & -1.700e+00 & 7 & 0.043 & 4420 & 26.82 & 3164 & 26.96 & 3767 & 26.87 & 3917 & 27.03 & 3393 & 27.43& 4.65\\
1442-0212 & 2.205e+02 & -2.195e+00 & 2 & 0.053 & 1605 & 26.35 & 1155 & 26.46 & 1355 & 26.40 & 1405 & 26.46 & 1252 & 26.88& 4.64\\
1503+3645 & 2.258e+02 & 3.674e+01 & 3 & 0.015 & 2208 & 26.66 & 1605 & 26.80 & 1908 & 26.69 & 2008 & 26.81 & 1443 & 27.25& 4.64\\
1519-0746 & 2.298e+02 & -7.767e+00 & 4 & 0.100 & 2511 & 26.34 & 1858 & 26.64 & 2208 & 26.49 & 2361 & 26.51 & 2244 & 26.91& 4.63\\
1520-2501 & 2.301e+02 & -2.502e+01 & 2 & 0.165 & 1505 & 26.23 & 1005 & 26.28 & 1255 & 26.29 & 1355 & 26.42 & 1872 & 27.01& 4.62\\
1524+0956 & 2.310e+02 & 9.939e+00 & 5 & 0.042 & 3364 & 26.74 & 2461 & 26.85 & 2914 & 26.88 & 3011 & 26.85 & 2520 & 27.31& 4.64\\
1525+0955 & 2.312e+02 & 9.921e+00 & 3 & 0.039 & 1608 & 26.14 & 1155 & 26.40 & 1355 & 26.40 & 1455 & 26.53 & 1230 & 26.87& 4.61\\
1525+0960 & 2.312e+02 & 1.000e+01 & 4 & 0.038 & 2861 & 26.68 & 2008 & 26.87 & 2361 & 26.74 & 2461 & 26.92 & 2154 & 27.22& 4.64\\
1536+1410 & 2.341e+02 & 1.417e+01 & 3 & 0.047 & 2408 & 26.62 & 1758 & 26.62 & 2108 & 26.74 & 2158 & 26.69 & 1950 & 27.19& 4.63\\
1558+0812 & 2.396e+02 & 8.196e+00 & 3 & 0.039 & 2308 & 26.60 & 1255 & 26.64 & 955 & 26.27 & 1505 & 26.58 & 1457 & 27.10& 4.64
\enddata
\label{tab:field}
\end{deluxetable*}

\begin{deluxetable*}{lrrccccccccccccc}
\tablenum{5}
\tabletypesize{\footnotesize}
\tablecolumns{14}
\tablewidth{0pt} 
\tablecaption{Continued.}
\tablehead{\colhead{Field ID} & \colhead{$\alpha$(J2000)} & \colhead{$\delta$(J2000)} & \colhead{No. of}  & \colhead{E(B-V)}  & \multicolumn{2}{c}{F350LP} & \multicolumn{2}{c}{F105W} &\multicolumn{2}{c}{F125W} & \multicolumn{2}{c}{F140W} & \multicolumn{2}{c}{F160W} & \colhead{$A_{\rm e}$}\\
        \colhead{}& \colhead{(deg)} & \colhead{(deg)}& \colhead{orbits} && \colhead{t$_{\rm exp}$}  & \colhead{m$_{\rm lim}$} & \colhead{t$_{\rm exp}$} & \colhead{m$_{\rm lim}$} & \colhead{t$_{\rm exp}$} & \colhead{m$_{\rm lim}$} & \colhead{t$_{\rm exp}$} & \colhead{m$_{\rm lim}$} & \colhead{t$_{\rm exp}$} & \colhead{m$_{\rm lim}$} & \colhead{($\square \arcmin$)}}
\startdata
1607+1332 & 2.417e+02 & 1.354e+01 & 3 & 0.036 & 2408 & 26.60 & 1758 & 26.76 & 2108 & 26.77 & 2158 & 26.76 & 1975 & 27.13& 4.64\\
1614+4856 & 2.435e+02 & 4.894e+01 & 4 & 0.014 & 3211 & 26.71 & 2308 & 26.97 & 2661 & 26.76 & 2711 & 26.75 & 2222 & 26.94& 4.63\\
1619+2541 & 2.448e+02 & 2.568e+01 & 3 & 0.048 & 2408 & 26.63 & 1758 & 26.77 & 2108 & 26.72 & 2158 & 26.72 & 1997 & 27.31& 4.63\\
1632+3736 & 2.479e+02 & 3.761e+01 & 12 & 0.010 & 8435 & 27.16 & 6026 & 27.40 & 7235 & 27.24 & 7385 & 27.34 & 6480 & 27.82& 5.28\\
1659+3732 & 2.548e+02 & 3.753e+01 & 3 & 0.017 & 2508 & 26.60 & 1808 & 26.76 & 2108 & 26.85 & 2158 & 26.71 & 2179 & 27.38& 4.62\\
1708+4237 & 2.571e+02 & 4.262e+01 & 3 & 0.024 & 2708 & 26.84 & 2108 & 26.90 & 2408 & 26.88 & 2408 & 26.86 & 2055 & 27.38& 4.64\\
1715+0455 & 2.588e+02 & 4.915e+00 & 4 & 0.119 & 2461 & 26.47 & 1908 & 26.70 & 2261 & 26.66 & 2411 & 26.66 & 933 & 26.66& 4.63\\
1715+0502 & 2.588e+02 & 5.035e+00 & 5 & 0.132 & 3064 & 26.63 & 2261 & 26.82 & 2661 & 26.71 & 2911 & 26.85 & 3262 & 27.41& 4.64\\
1738+1839 & 2.644e+02 & 1.865e+01 & 2 & 0.061 & 1605 & 26.40 & 1155 & 26.51 & 1355 & 26.52 & 1505 & 26.32 & 1541 & 27.07& 4.64\\
1920-4531 & 2.901e+02 & -4.552e+01 & 4 & 0.086 & 2911 & 26.51 & 2058 & 26.74 & 2511 & 26.35 & 2661 & 26.48 & 2660 & 26.98& 4.62\\
2008-6610 & 3.020e+02 & -6.617e+01 & 2 & 0.071 & 1805 & 26.44 & 1305 & 26.63 & 1505 & 26.49 & 1605 & 26.07 & 1855 & 27.08& 4.64\\
2057-1423 & 3.143e+02 & -1.438e+01 & 3 & 0.051 & 1908 & 26.32 & 1505 & 26.53 & 1705 & 26.45 & 1705 & 26.44 & 1710 & 27.10& 4.64\\
2134-0708 & 3.235e+02 & -7.126e+00 & 6 & 0.033 & 4167 & 26.28 & 2964 & 26.54 & 3514 & 26.35 & 3714 & 26.27 & 3605 & 26.92& 4.70\\
2140+0241 & 3.249e+02 & 2.690e+00 & 2 & 0.088 & 1605 & 26.37 & 1155 & 26.45 & 1355 & 26.31 & 1405 & 26.30 & 1872 & 27.09& 4.61\\
2141-2310$^\ast$ & 3.252e+02 & -2.317e+01 & 3 & 0.049 & --- & --- & --- & --- & --- & --- & --- & --- & --- & --- & ---\\
2228-0955 & 3.371e+02 & -9.920e+00 & 3 & 0.052 & 2408 & 26.38 & 1708 & 26.52 & 2058 & 26.47 & 2108 & 26.50 & 2093 & 26.97& 4.64\\
2229-0945 & 3.372e+02 & -9.752e+00 & 3 & 0.050 & 1758 & 26.26 & 1255 & 26.39 & 1505 & 26.40 & 1605 & 26.32 & 1479 & 26.83& 4.63\\
2253-1411 & 3.434e+02 & -1.419e+01 & 4 & 0.044 & 2761 & 26.57 & 2008 & 26.70 & 2361 & 26.65 & 2461 & 26.75 & 2174 & 27.09& 4.61\\
2312-1423 & 3.479e+02 & -1.439e+01 & 3 & 0.035 & 2408 & 26.55 & 1758 & 26.57 & 2108 & 26.72 & 2158 & 26.57 & 1970 & 27.09& 4.64\\
2323+0059 & 3.507e+02 & -9.841e-01 & 3 & 0.044 & 2408 & 26.37 & 1708 & 26.42 & 2008 & 26.34 & 2158 & 26.40 & 2095 & 26.94& 4.63
\enddata
\tablecomments{NOTE: 
Column 1: Field name derived from the coordinates. 
Columns 2-3: $\alpha$ and $\delta$ coordinates (in degrees) as from the F140W exposure. 
Column 4: total number of {\it HST} orbits allocated. 
Column 5: Galactic extinction E(B-V) from \cite{schlafly11}. 
Columns 6-15: exposure time (in seconds) and $5\sigma$ limiting magnitude (in AB magnitudes) within a $r=0\farcs32$ aperture in each band. 
Column 16: effective area (in arcmin$^2$).}
\tablenotetext{$^\ast$}{Field excessively dominated by stars.}
\tablenotetext{$^\dagger$}{Guide star acquisition failure.}
\end{deluxetable*}

\section{B. Rejected candidates}\label{sec:Ab}
We summarize here objects that are selected in the color-color selection (Section \ref{ssec:cs}) but rejected in the later selection processes.
Out of \njjcol\ (\nyycol) objects selected with the color selection as $z\sim10$ ($z\sim9$) candidates, 0 (\nyyoap) are rejected by the small aperture photometry in \oo\ (Section~\ref{ssec:cs}), \njjvis\ (\nyyvis) by the visual inspection (Section~\ref{ssec:vis}), \njjpzori\ (\nyypzori) by photometric redshifts without priors (Section~\ref{sssec:fp}), \njjpz\ (\nyypz) by photometric redshifts with priors (Section~\ref{sssec:mp}), and \njhlv\ (\nyylv) with additional F098M photometry (Section~\ref{ssec:f098m}). 
The rejected objects are summarized in Table~\ref{tab:rej}.

\begin{figure*}
\centering
\includegraphics[width=0.7\textwidth]{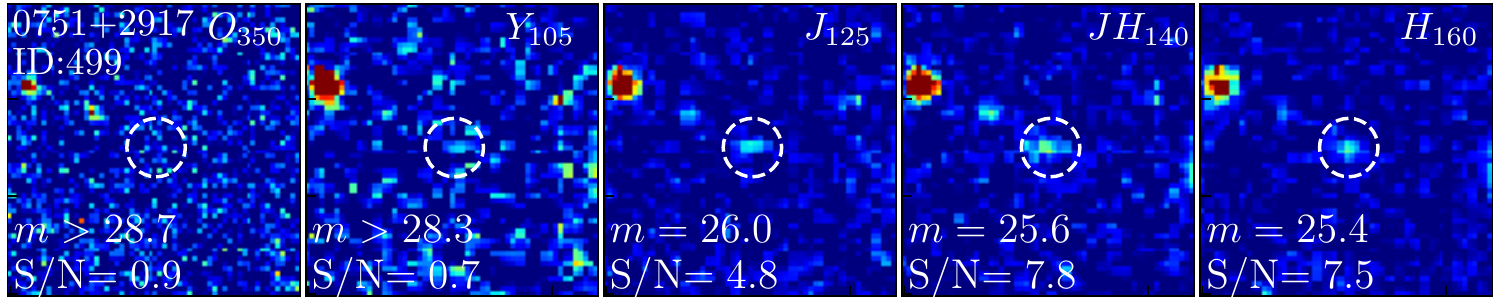}
\includegraphics[width=0.14\textwidth]{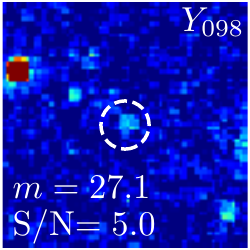}
\includegraphics[width=0.8\textwidth]{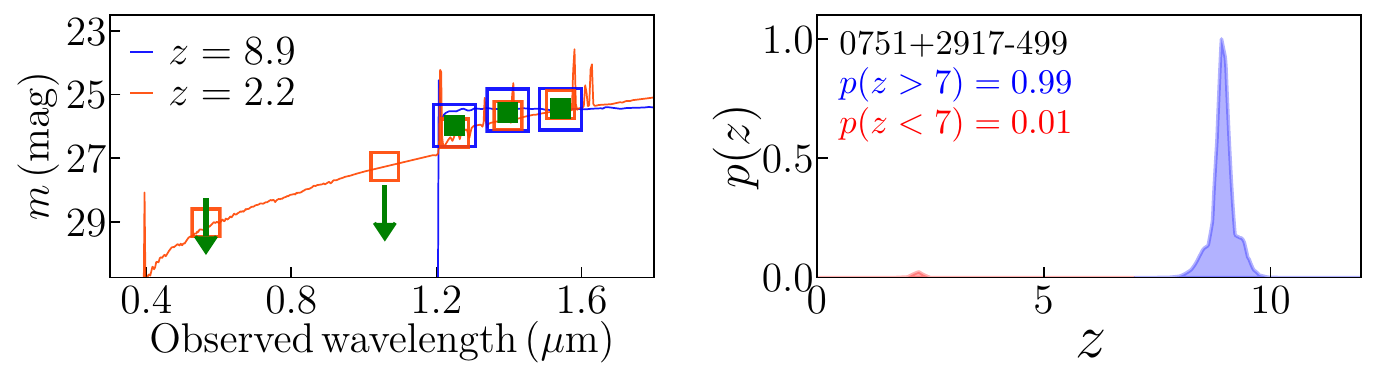}
\includegraphics[width=0.8\textwidth]{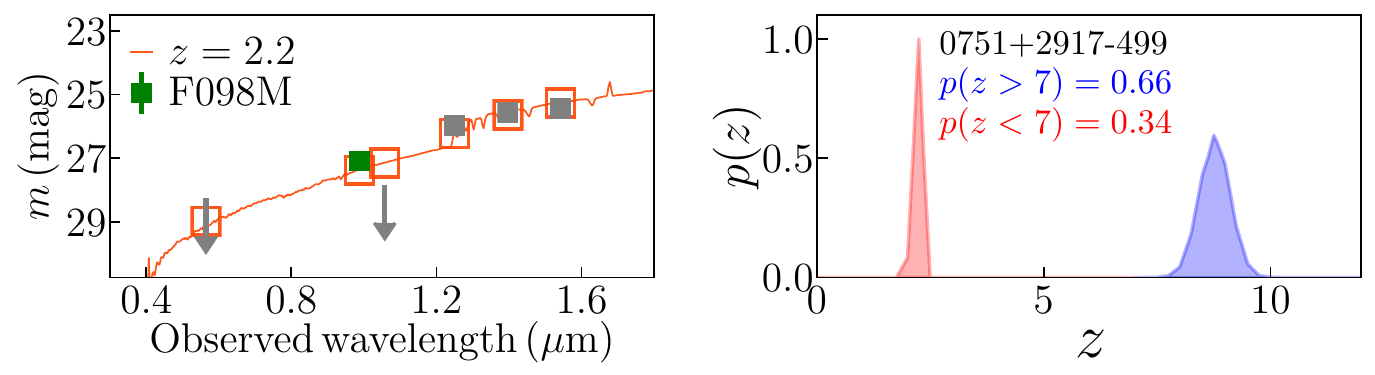}
\caption{
Same as Fig.~\ref{fig:jh} but for 0751+2917-499 with F098M detection (top right), despite the absence of clear detection in F105W.
The color stretch is changed for $Y_{105}$ from other filters to stress its noise level. 
The central blob is hardly distinguishable from the surrounding noise either by eyes or SExtractor ($S/N=0.7$). 
The non-detection in \yy\ is attributed to its shallow exposure ($t=2008$\,sec), while F098M has a much deep exposure from multiple observing programs ($t=18642$\,sec).
Addition of F098M photometry dramatically changes the photometric probability distribution (from middle to bottom panels), revealing the importance of medium band photometry near Lyman breaks.
}
\label{fig:f098m}
\end{figure*}

{\section{C. Importance of supplemental F098M photometry}\label{sec:Ac}
Two among the final $z\sim9$ candidates were rejected after adding supplemental F098M photometry (one in this study and one in \citealt{livermore18}). These candidates would otherwise have been included in the final sample, satisfying all the criteria in this study (i.e. $<20\%$ of contamination rate).
In particular, without F098M data, the redshift probability distribution for the former source included a $>98\%$ likelihood at $z>8$.
With the F098M photometry ($m_{098}=27.1\pm0.2$), the redshift probability changed dramatically, preferring a low-$z$ result at $z\sim2.2$ (Fig.~\ref{fig:f098m}).

With F098M, the SED (Fig.~\ref{fig:f098m}) is now fitted with a passive galaxy template, whose best-fit parameters are $\log t/{\rm yr}=8.6_{-0.6}^{+0.4}$, $A_V/{\rm mag}=0.8_{-0.8}^{+1.9}$, and $\logm=9.7_{-0.3}^{+0.6}$, revealing this as an intermediate-mass, moderately dusty, quiescent galaxy.
The SED parameters are derived from FAST \citep{kriek09}, assuming the dust law of \citet{calzetti00} and the initial mass function of \citet{salpeter55}.
The best-fit template indicates the magnitude in $Y_{105}$ to be $\sim27$\,mag.

Our results and the recent work by \citet{livermore18} jointly
highlight the benefit from additional follow-up data with medium-band
filters around the expected Lyman break in order to build reliable
photometric samples. This approach is more efficient than just
application of a yet stricter selection criterion (which risks
decreasing significantly the selection completeness), and would
require only a modest amount of additional observing time because $z>8$
candidates are rare.}

\begin{deluxetable*}{lrrccccccccccccc}
\tabletypesize{\footnotesize}
\tablecolumns{16}
\tablewidth{0pt} 
\tablecaption{Coordinates and photometric properties of rejected objects}
\tablehead{\colhead{Obj ID} & \colhead{$\alpha$(J2000)} & \colhead{$\delta$(J2000)} & \colhead{\hh} & \multicolumn{2}{c}{Colors} & \multicolumn{6}{c}{$S/N$} & \multicolumn{2}{c}{$z_{\rm phot.}$} & \colhead{$p_{\rm l}/p_{\rm h}$} & \colhead{Reason}\\
\colhead{} & \colhead{(deg)} & \colhead{(deg)}& \colhead{(mag)} &\colhead{$C_1$} & \colhead{$C_2$}& \colhead{$O_{\rm ap.}$}& \colhead{\oo} & \colhead{\yy} & \colhead{\jj} & \colhead{\jh} & \colhead{\hh} & \colhead{peak} & \colhead{prior} & \colhead{} & \colhead{of rejection}}
\startdata
\cutinhead{$z\sim10$ candidates}
0807+3606-406 & 121.854693 & 36.100561 & $25.4$ & $1.3\pm0.3$ & $0.6\pm0.2$ & $-0.3$ & $-3.2$ & $0.9$ & $3.3$ & $6.6$ & $9.3$ & 9.3 & 3.1 & 2.28  & $z_\mathrm{prior}$\\
0834+5238-575 & 128.508154 & 52.632111 & $26.3$ & $2.0\pm0.8$ & $0.9\pm0.3$ & $0.5$ & $0.4$ & $0.9$ & $1.3$ & $4.0$ & $7.5$ & 7.0 & 3.6 & 2.16  & vis\\
0851+4240-465 & 132.729208 & 42.670074 & $24.8$ & $1.7\pm0.5$ & $0.6\pm0.2$ & $1.0$ & $1.2$ & $0.9$ & $2.2$ & $6.5$ & $10.5$ & 10.1 & 3.4 & 0.95  & $z_\mathrm{prior}$\\
0948+5757-185 & 147.048362 & 57.949351 & $25.4$ & $2.1\pm0.9$ & $0.8\pm0.2$ & $0.4$ & $-0.0$ & $1.3$ & $1.2$ & $5.8$ & $10.0$ & 7.2 & 3.9 & 0.53  & $z_\mathrm{prior}$\\
0953+5153-1073 & 148.299729 & 51.877393 & $26.1$ & $1.4\pm0.6$ & $0.6\pm0.3$ & $1.1$ & $1.0$ & $1.0$ & $2.0$ & $4.5$ & $6.5$ & 7.0 & 2.8 & 5.45  & $z_\mathrm{prior}$\\
0956+2848-401 & 149.093549 & 28.792157 & $26.2$ & $1.4\pm0.5$ & $0.7\pm0.3$ & $1.0$ & $1.0$ & $0.9$ & $2.1$ & $4.5$ & $7.6$ & 9.9 & 3.1 & 1.41  & $z_\mathrm{prior}$\\
1136+0747-805 & 173.930982 & 7.777363 & $25.8$ & $2.2\pm1.1$ & $0.8\pm0.3$ & $0.7$ & $1.4$ & $0.8$ & $1.0$ & $3.5$ & $7.3$ & 10.3 & 2.8 & 1.09  & vis\\
1152+5433-109 & 177.928053 & 54.574866 & $25.8$ & $1.4\pm0.5$ & $0.6\pm0.2$ & $-0.1$ & $1.1$ & $1.3$ & $2.2$ & $5.5$ & $7.4$ & 7.0 & 3.0 & 2.55  & $z_\mathrm{prior}$\\
1218+3008-265 & 184.563250 & 30.136829 & $24.4$ & $1.8\pm0.2$ & $0.7\pm0.1$ & $-0.3$ & $0.2$ & $0.7$ & $4.4$ & $13.1$ & $21.4$ & 10.1 & 3.5 & 0.47  & $z_\mathrm{prior}$\\
1619+2541-760 & 244.807436 & 25.686392 & $26.5$ & $2.2\pm1.0$ & $0.9\pm0.3$ & $-1.4$ & $-1.0$ & $1.0$ & $1.1$ & $3.5$ & $7.3$ & 7.5 & 3.9 & 0.11  & vis\\
1738+1839-581 & 264.400431 & 18.664927 & $27.0$ & $1.5\pm0.7$ & $0.6\pm0.3$ & $0.7$ & $1.3$ & $-0.3$ & $1.7$ & $3.8$ & $6.1$ & 9.9 & 3.2 & 1.49  & vis\\
2134-0708-645$\dagger$ & 323.562304 & -7.120044 & $25.4$ & $1.8\pm0.8$ & $0.5\pm0.2$ & $0.5$ & $1.0$ & $-0.2$ & $1.4$ & $5.7$ & $6.9$ & 10.0 & 10.4 & 0.12  & $z_\mathrm{prior}$\\
2229-0945-424 & 337.196149 & -9.738446 & $26.4$ & $1.8\pm0.6$ & $0.8\pm0.3$ & $-1.1$ & $-0.4$ & $-0.2$ & $1.8$ & $4.8$ & $8.5$ & 10.2 & 3.5 & 6.05  & vis\\
2312-1423-353 & 347.931939 & -14.404852 & $25.4$ & $1.7\pm0.5$ & $1.0\pm0.3$ & $-0.2$ & $-0.4$ & $0.5$ & $2.1$ & $3.6$ & $9.1$ & 10.2 & 3.6 & 1.75  & $z_\mathrm{prior}$\\

\cutinhead{$z\sim9$ candidates}
0116+1425-442 & 19.034665 & 14.402626 & $24.4$ & $1.8\pm0.4$ & $0.1\pm0.1$ & $0.2$ & $-0.8$ & $2.7$ & $11.0$ & $14.4$ & $16.4$ & 8.4 & 8.7 & 0.32  & $Y_{098}$\\
0314-6712-1393 & 48.393390 & -67.198000 & $26.5$ & $1.8\pm0.8$ & $0.0\pm0.3$ & $0.8$ & $0.7$ & $1.3$ & $3.6$ & $6.1$ & $5.6$ & 8.4 & 2.0 & 0.87  & $z_\mathrm{prior}$\\
0337-0507-775 & 54.365817 & -5.118763 & $25.0$ & $1.8\pm0.6$ & $-0.0\pm0.1$ & $3.0$ & $1.2$ & $1.9$ & $6.1$ & $13.1$ & $10.8$ & 7.7 & 2.0 & 1.05  & $O_{ap.}$\\
0554-6005-812 & 88.383324 & -60.080277 & $26.8$ & $>1.9$ & $-0.1\pm0.3$ & $0.7$ & $1.1$ & $0.1$ & $3.2$ & $6.8$ & $4.9$ & 9.2 & 2.1 & 0.38  & $z_\mathrm{prior}$\\
0751+2917-410 & 117.723475 & 29.287342 & $25.1$ & $>3.0$ & $-0.0\pm0.1$ & $0.0$ & $-4.7$ & $0.0$ & $11.9$ & $12.6$ & $10.7$ & 9.2 & 8.8 & 0.17  & vis\\
0751+2917-499 & 117.717853 & 29.288468 & $25.4$ & $>2.4$ & $0.1\pm0.2$ & $0.0$ & $0.9$ & $0.7$ & $4.8$ & $7.8$ & $7.5$ & 9.0 & 8.9 & 0.05  & $Y_{098}$\\
0952+5150-162 & 148.066856 & 51.817542 & $25.4$ & $1.6\pm0.5$ & $0.1\pm0.2$ & $0.6$ & $0.4$ & $2.3$ & $4.4$ & $9.4$ & $8.4$ & 7.7 & 2.0 & 0.79  & $z_\mathrm{prior}$\\
1017-2052-119 & 154.367086 & -20.858955 & $24.9$ & $1.6\pm0.6$ & $0.1\pm0.2$ & $0.1$ & $-0.6$ & $2.0$ & $4.4$ & $9.5$ & $9.0$ & 7.4 & 2.0 & 1.10  & $z_\mathrm{prior}$\\
1104+2813-156 & 165.968967 & 28.198873 & $24.9$ & $1.7\pm0.5$ & $0.0\pm0.1$ & $0.9$ & $-0.3$ & $2.3$ & $5.6$ & $11.5$ & $10.9$ & 7.5 & 2.0 & 1.07  & $z_\mathrm{prior}$\\
1152+5433-577 & 177.915777 & 54.541200 & $26.4$ & $>1.5$ & $0.1\pm0.3$ & $-0.4$ & $-0.4$ & $0.1$ & $3.9$ & $6.2$ & $5.0$ & 8.7 & 8.7 & 0.08  & vis\\
1152+5433-664 & 177.922295 & 54.541692 & $26.4$ & $>1.5$ & $-0.1\pm0.3$ & $0.3$ & $-0.1$ & $-3.5$ & $3.7$ & $6.1$ & $4.4$ & 9.0 & 9.0 & 0.01  & vis\\
1438-0142-939 & 219.459857 & -1.683573 & $26.6$ & $>2.8$ & $-0.9\pm0.3$ & $0.1$ & $-0.5$ & $0.3$ & $0.1$ & $10.5$ & $4.2$ & 10.5 & 2.0 & 1.14  & vis\\
1442-0212-600 & 220.555169 & -2.177198 & $25.6$ & $>2.2$ & $0.2\pm0.2$ & $0.9$ & $0.8$ & $0.9$ & $4.9$ & $7.9$ & $6.8$ & 8.8 & 8.8 & 0.22  & $z_\mathrm{flat}$\\
1520-2501-228 & 230.076333 & -25.013525 & $24.2$ & $1.7\pm0.3$ & $0.2\pm0.1$ & $0.6$ & $-0.1$ & $3.1$ & $8.0$ & $14.1$ & $16.4$ & 7.5 & 8.6 & 0.72  & $z_\mathrm{flat}$\\
2008-6610-77 & 302.024097 & -66.165779 & $25.7$ & $>2.4$ & $0.1\pm0.2$ & $0.2$ & $-0.6$ & $0.8$ & $8.7$ & $8.6$ & $8.4$ & 8.9 & 8.9 & 0.34  & $z_\mathrm{flat}$\\
2229-0945-755 & 337.194543 & -9.768108 & $26.3$ & $2.1\pm1.0$ & $-0.2\pm0.2$ & $0.7$ & $0.0$ & $1.1$ & $7.8$ & $8.9$ & $7.0$ & 8.8 & 8.8 & 0.02  & vis\\
2312-1423-763 & 347.926274 & -14.380445 & $26.1$ & $>2.1$ & $0.1\pm0.2$ & $-0.2$ & $1.0$ & $-0.8$ & $3.2$ & $6.6$ & $7.0$ & 10.1 & 8.8 & 0.61  & $z_\mathrm{prior}$\\

\enddata
\tablecomments{
Same as Table~\ref{tab:jh}, but for those color selected objects rejected by either of followings. 
Reason of rejection---vis: Visual inspection. $z_{\rm flat}$: Photometric redshifts with a flat prior show inconsistent results with color-cut criteria and/or the peak of high-$z$ probability is not sufficient compared to the low-$z$ one ($p_{\rm low}/p_{\rm high}>0.2$). $z_{\rm prior}$: Same as $z_{\rm flat}$ but for photometric redshifts with a custom prior (Section~\ref{sssec:ap}). $O_{ap.}$: Excess of $O_{350}$ flux (S/N\,$>1.5$) in a small aperture (radius of 0.\!\arcsec16). $Y_{098}$: Follow-up study rejects the high-$z$ possibility \citep{livermore18}. 
$\dagger:$ Candidates selected in C16.
}
\label{tab:rej}
\end{deluxetable*}


\clearpage

\bibliographystyle{apj}
\bibliography{/Users/tmorishita/adssample.bib}

\begin{thebibliography}{}
\expandafter\ifx\csname natexlab\endcsname\relax\def\natexlab#1{#1}\fi

\bibitem[{{Anderson}(2014)}]{anderson14}
{Anderson}, J. 2014, {The Impact of x-CTE in the WFC3/UVIS detector on
  Astrometry}, Tech. rep.

\bibitem[{{Atek} {et~al.}(2018){Atek}, {Richard}, {Kneib}, \&
  {Schaerer}}]{atek18}
{Atek}, H., {Richard}, J., {Kneib}, J.-P., \& {Schaerer}, D. 2018, ArXiv
  e-prints, arXiv:1803.09747

\bibitem[{{Atek} {et~al.}(2015){Atek}, {Richard}, {Jauzac}, {Kneib},
  {Natarajan}, {Limousin}, {Schaerer}, {Jullo}, {Ebeling}, {Egami}, \&
  {Clement}}]{atek15}
{Atek}, H., {Richard}, J., {Jauzac}, M., {et~al.} 2015, \apj, 814,
  doi:10.1088/0004-637X/814/1/69

\bibitem[{{Barone-Nugent} {et~al.}(2014){Barone-Nugent}, {Trenti}, {Wyithe},
  {Bouwens}, {Oesch}, {Illingworth}, {Carollo}, {Su}, {Stiavelli}, {Labbe}, \&
  {van Dokkum}}]{baronenugent14}
{Barone-Nugent}, R.~L., {Trenti}, M., {Wyithe}, J.~S.~B., {et~al.} 2014, \apj,
  793, 17

\bibitem[{{Behroozi} {et~al.}(2013){Behroozi}, {Wechsler}, \&
  {Conroy}}]{behroozi13}
{Behroozi}, P.~S., {Wechsler}, R.~H., \& {Conroy}, C. 2013, \apj, 770, 57

\bibitem[{{Ben{\'{\i}}tez}(2000)}]{benitez00}
{Ben{\'{\i}}tez}, N. 2000, \apj, 536, 571

\bibitem[{{Bernard} {et~al.}(2016){Bernard}, {Carrasco}, {Trenti}, {Oesch},
  {Wu}, {Bradley}, {Schmidt}, {Bouwens}, {Calvi}, {Mason}, {Stiavelli}, \&
  {Treu}}]{bernard16}
{Bernard}, S.~R., {Carrasco}, D., {Trenti}, M., {et~al.} 2016, \apj, 827, 76

\bibitem[{{Bertin} \& {Arnouts}(1996)}]{bertin96}
{Bertin}, E., \& {Arnouts}, S. 1996, \aaps, 117, 393

\bibitem[{{Birnboim} \& {Dekel}(2003)}]{birnboim03}
{Birnboim}, Y., \& {Dekel}, A. 2003, \mnras, 345, 349

\bibitem[{{Bouwens} {et~al.}(2017){Bouwens}, {Illingworth}, {Oesch}, {Atek},
  {Lam}, \& {Stefanon}}]{bouwens17}
{Bouwens}, R.~J., {Illingworth}, G.~D., {Oesch}, P.~A., {et~al.} 2017, \apj,
  843, 41

\bibitem[{{Bouwens} {et~al.}(2015{\natexlab{a}}){Bouwens}, {Illingworth},
  {Oesch}, {Caruana}, {Holwerda}, {Smit}, \& {Wilkins}}]{bouwens15b}
---. 2015{\natexlab{a}}, \apj, 811, 140

\bibitem[{{Bouwens} {et~al.}(2014){Bouwens}, {Illingworth}, {Oesch},
  {Labb{\'e}}, {van Dokkum}, {Trenti}, {Franx}, {Smit}, {Gonzalez}, \&
  {Magee}}]{bouwens14}
---. 2014, \apj, 793, 115

\bibitem[{{Bouwens} {et~al.}(2015{\natexlab{b}}){Bouwens}, {Illingworth},
  {Oesch}, {Trenti}, {Labb{\'e}}, {Bradley}, {Carollo}, {van Dokkum},
  {Gonzalez}, {Holwerda}, {Franx}, {Spitler}, {Smit}, \& {Magee}}]{bouwens15}
---. 2015{\natexlab{b}}, \apj, 803, 34

\bibitem[{{Bouwens} {et~al.}(2016){Bouwens}, {Aravena}, {Decarli}, {Walter},
  {da Cunha}, {Labb{\'e}}, {Bauer}, {Bertoldi}, {Carilli}, {Chapman}, {Daddi},
  {Hodge}, {Ivison}, {Karim}, {Le Fevre}, {Magnelli}, {Ota}, {Riechers},
  {Smail}, {van der Werf}, {Weiss}, {Cox}, {Elbaz}, {Gonzalez-Lopez},
  {Infante}, {Oesch}, {Wagg}, \& {Wilkins}}]{bouwens16}
{Bouwens}, R.~J., {Aravena}, M., {Decarli}, R., {et~al.} 2016, \apj, 833, 72

\bibitem[{{Bradley} {et~al.}(2012){Bradley}, {Trenti}, {Oesch}, {Stiavelli},
  {Treu}, {Bouwens}, {Shull}, {Holwerda}, \& {Pirzkal}}]{bradley12}
{Bradley}, L.~D., {Trenti}, M., {Oesch}, P.~A., {et~al.} 2012, \apj, 760, 108

\bibitem[{{Brammer} {et~al.}(2014){Brammer}, {Pirzkal}, {McCullough}, \&
  {MacKenty}}]{brammer14}
{Brammer}, G., {Pirzkal}, N., {McCullough}, P., \& {MacKenty}, J. 2014,
  {Time-varying Excess Earth-glow Backgrounds in the WFC3/IR Channel}, Tech.
  rep.

\bibitem[{{Brammer} {et~al.}(2008){Brammer}, {van Dokkum}, \&
  {Coppi}}]{brammer08}
{Brammer}, G.~B., {van Dokkum}, P.~G., \& {Coppi}, P. 2008, \apj, 686, 1503

\bibitem[{{Brinchmann} {et~al.}(2017){Brinchmann}, {Inami}, {Bacon}, {Contini},
  {Maseda}, {Chevallard}, {Bouch{\'e}}, {Boogaard}, {Carollo}, {Charlot},
  {Kollatschny}, {Marino}, {Pello}, {Richard}, {Schaye}, {Verhamme}, \&
  {Wisotzki}}]{brinchmann17}
{Brinchmann}, J., {Inami}, H., {Bacon}, R., {et~al.} 2017, \aap, 608, A3

\bibitem[{{Bruzual} \& {Charlot}(2003)}]{bruzual03}
{Bruzual}, G., \& {Charlot}, S. 2003, \mnras, 344, 1000

\bibitem[{{Calvi} {et~al.}(2016){Calvi}, {Trenti}, {Stiavelli}, {Oesch},
  {Bradley}, {Schmidt}, {Coe}, {Brammer}, {Bernard}, {Bouwens}, {Carrasco},
  {Carollo}, {Holwerda}, {MacKenty}, {Mason}, {Shull}, \& {Treu}}]{calvi16}
{Calvi}, V., {Trenti}, M., {Stiavelli}, M., {et~al.} 2016, \apj, 817, 120

\bibitem[{{Calzetti} {et~al.}(2000){Calzetti}, {Armus}, {Bohlin}, {Kinney},
  {Koornneef}, \& {Storchi-Bergmann}}]{calzetti00}
{Calzetti}, D., {Armus}, L., {Bohlin}, R.~C., {et~al.} 2000, \apj, 533, 682

\bibitem[{{Cardelli} {et~al.}(1989){Cardelli}, {Clayton}, \&
  {Mathis}}]{cardelli89}
{Cardelli}, J.~A., {Clayton}, G.~C., \& {Mathis}, J.~S. 1989, \apj, 345, 245

\bibitem[{{Carrasco} {et~al.}(2018){Carrasco}, {Trenti}, {Mutch}, \&
  {Oesch}}]{carrasco18}
{Carrasco}, D., {Trenti}, M., {Mutch}, S., \& {Oesch}, P. 2018, ArXiv e-prints,
  arXiv:1805.08985

\bibitem[{{Casertano} {et~al.}(2000){Casertano}, {de Mello}, {Dickinson},
  {Ferguson}, {Fruchter}, {Gonzalez-Lopezlira}, {Heyer}, {Hook}, {Levay},
  {Lucas}, {Mack}, {Makidon}, {Mutchler}, {Smith}, {Stiavelli}, {Wiggs}, \&
  {Williams}}]{casertano00}
{Casertano}, S., {de Mello}, D., {Dickinson}, M., {et~al.} 2000, \aj, 120, 2747

\bibitem[{{Clay} {et~al.}(2015){Clay}, {Thomas}, {Wilkins}, \&
  {Henriques}}]{clay15}
{Clay}, S.~J., {Thomas}, P.~A., {Wilkins}, S.~M., \& {Henriques}, B.~M.~B.
  2015, \mnras, 451, 2692

\bibitem[{{Coe} {et~al.}(2006){Coe}, {Ben{\'{\i}}tez}, {S{\'a}nchez}, {Jee},
  {Bouwens}, \& {Ford}}]{coe06}
{Coe}, D., {Ben{\'{\i}}tez}, N., {S{\'a}nchez}, S.~F., {et~al.} 2006, \aj, 132,
  926

\bibitem[{{Coe} {et~al.}(2013){Coe}, {Zitrin}, {Carrasco}, {Shu}, {Zheng},
  {Postman}, {Bradley}, {Koekemoer}, {Bouwens}, {Broadhurst}, {Monna}, {Host},
  {Moustakas}, {Ford}, {Moustakas}, {van der Wel}, {Donahue}, {Rodney},
  {Ben{\'{\i}}tez}, {Jouvel}, {Seitz}, {Kelson}, \& {Rosati}}]{coe13}
{Coe}, D., {Zitrin}, A., {Carrasco}, M., {et~al.} 2013, \apj, 762, 32

\bibitem[{{Cowley} {et~al.}(2018){Cowley}, {Baugh}, {Cole}, {Frenk}, \&
  {Lacey}}]{cowley18}
{Cowley}, W.~I., {Baugh}, C.~M., {Cole}, S., {Frenk}, C.~S., \& {Lacey}, C.~G.
  2018, \mnras, 474, 2352

\bibitem[{{Curtis-Lake} {et~al.}(2016){Curtis-Lake}, {McLure}, {Dunlop},
  {Rogers}, {Targett}, {Dekel}, {Ellis}, {Faber}, {Ferguson}, {Grogin},
  {Kocevski}, {Koekemoer}, {Lai}, {M{\'a}rmol-Queralt{\'o}}, \&
  {Robertson}}]{curtis-Lake16}
{Curtis-Lake}, E., {McLure}, R.~J., {Dunlop}, J.~S., {et~al.} 2016, \mnras,
  457, 440

\bibitem[{{Ellis} {et~al.}(2013){Ellis}, {McLure}, {Dunlop}, {Robertson},
  {Ono}, {Schenker}, {Koekemoer}, {Bowler}, {Ouchi}, {Rogers}, {Curtis-Lake},
  {Schneider}, {Charlot}, {Stark}, {Furlanetto}, \& {Cirasuolo}}]{ellis13}
{Ellis}, R.~S., {McLure}, R.~J., {Dunlop}, J.~S., {et~al.} 2013, \apjl, 763, L7

\bibitem[{{Erb} {et~al.}(2010){Erb}, {Pettini}, {Shapley}, {Steidel}, {Law}, \&
  {Reddy}}]{erb10}
{Erb}, D.~K., {Pettini}, M., {Shapley}, A.~E., {et~al.} 2010, \apj, 719, 1168

\bibitem[{{Faber} \& {Jackson}(1976)}]{faber76}
{Faber}, S.~M., \& {Jackson}, R.~E. 1976, \apj, 204, 668

\bibitem[{{Fan} {et~al.}(2006){Fan}, {Carilli}, \& {Keating}}]{fan06}
{Fan}, X., {Carilli}, C.~L., \& {Keating}, B. 2006, \araa, 44, 415

\bibitem[{{Fan} {et~al.}(2000){Fan}, {White}, {Davis}, {Becker}, {Strauss},
  {Haiman}, {Schneider}, {Gregg}, {Gunn}, {Knapp}, {Lupton}, {Anderson},
  {Anderson}, {Annis}, {Bahcall}, {Boroski}, {Brunner}, {Chen}, {Connolly},
  {Csabai}, {Doi}, {Fukugita}, {Hennessy}, {Hindsley}, {Ichikawa},
  {Ivezi{\'c}}, {Loveday}, {Meiksin}, {McKay}, {Munn}, {Newberg}, {Nichol},
  {Okamura}, {Pier}, {Sekiguchi}, {Shimasaku}, {Stoughton}, {Szalay},
  {Szokoly}, {Thakar}, {Vogeley}, \& {York}}]{fan00}
{Fan}, X., {White}, R.~L., {Davis}, M., {et~al.} 2000, \aj, 120, 1167

\bibitem[{{Fan} {et~al.}(2003){Fan}, {Strauss}, {Schneider}, {Becker}, {White},
  {Haiman}, {Gregg}, {Pentericci}, {Grebel}, {Narayanan}, {Loh}, {Richards},
  {Gunn}, {Lupton}, {Knapp}, {Ivezi{\'c}}, {Brandt}, {Collinge}, {Hao},
  {Harbeck}, {Prada}, {Schaye}, {Strateva}, {Zakamska}, {Anderson},
  {Brinkmann}, {Bahcall}, {Lamb}, {Okamura}, {Szalay}, \& {York}}]{fan03}
{Fan}, X., {Strauss}, M.~A., {Schneider}, D.~P., {et~al.} 2003, \aj, 125, 1649

\bibitem[{{Finkelstein} {et~al.}(2015){Finkelstein}, {Ryan}, {Papovich},
  {Dickinson}, {Song}, {Somerville}, {Ferguson}, {Salmon}, {Giavalisco},
  {Koekemoer}, {Ashby}, {Behroozi}, {Castellano}, {Dunlop}, {Faber}, {Fazio},
  {Fontana}, {Grogin}, {Hathi}, {Jaacks}, {Kocevski}, {Livermore}, {McLure},
  {Merlin}, {Mobasher}, {Newman}, {Rafelski}, {Tilvi}, \&
  {Willner}}]{finkelstein15}
{Finkelstein}, S.~L., {Ryan}, Jr., R.~E., {Papovich}, C., {et~al.} 2015, \apj,
  810, 71

\bibitem[{{Finlator} {et~al.}(2011){Finlator}, {Oppenheimer}, \&
  {Dav{\'e}}}]{finlator11}
{Finlator}, K., {Oppenheimer}, B.~D., \& {Dav{\'e}}, R. 2011, \mnras, 410, 1703

\bibitem[{{Foreman-Mackey} {et~al.}(2013){Foreman-Mackey}, {Hogg}, {Lang}, \&
  {Goodman}}]{foreman13}
{Foreman-Mackey}, D., {Hogg}, D.~W., {Lang}, D., \& {Goodman}, J. 2013, \pasp,
  125, 306

\bibitem[{{Gehrels}(1986)}]{gehrels86}
{Gehrels}, N. 1986, \apj, 303, 336

\bibitem[{{Grazian} {et~al.}(2012){Grazian}, {Castellano}, {Fontana},
  {Pentericci}, {Dunlop}, {McLure}, {Koekemoer}, {Dickinson}, {Faber},
  {Ferguson}, {Galametz}, {Giavalisco}, {Grogin}, {Hathi}, {Kocevski}, {Lai},
  {Newman}, \& {Vanzella}}]{grazian12}
{Grazian}, A., {Castellano}, M., {Fontana}, A., {et~al.} 2012, \aap, 547, A51

\bibitem[{{Grogin} {et~al.}(2011){Grogin}, {Kocevski}, {Faber}, {Ferguson},
  {Koekemoer}, {Riess}, {Acquaviva}, {Alexander}, {Almaini}, {Ashby}, {Barden},
  {Bell}, {Bournaud}, {Brown}, {Caputi}, {Casertano}, {Cassata}, {Castellano},
  {Challis}, {Chary}, {Cheung}, {Cirasuolo}, {Conselice}, {Roshan Cooray},
  {Croton}, {Daddi}, {Dahlen}, {Dav{\'e}}, {de Mello}, {Dekel}, {Dickinson},
  {Dolch}, {Donley}, {Dunlop}, {Dutton}, {Elbaz}, {Fazio}, {Filippenko},
  {Finkelstein}, {Fontana}, {Gardner}, {Garnavich}, {Gawiser}, {Giavalisco},
  {Grazian}, {Guo}, {Hathi}, {H{\"a}ussler}, {Hopkins}, {Huang}, {Huang},
  {Jha}, {Kartaltepe}, {Kirshner}, {Koo}, {Lai}, {Lee}, {Li}, {Lotz}, {Lucas},
  {Madau}, {McCarthy}, {McGrath}, {McIntosh}, {McLure}, {Mobasher},
  {Moustakas}, {Mozena}, {Nandra}, {Newman}, {Niemi}, {Noeske}, {Papovich},
  {Pentericci}, {Pope}, {Primack}, {Rajan}, {Ravindranath}, {Reddy}, {Renzini},
  {Rix}, {Robaina}, {Rodney}, {Rosario}, {Rosati}, {Salimbeni}, {Scarlata},
  {Siana}, {Simard}, {Smidt}, {Somerville}, {Spinrad}, {Straughn}, {Strolger},
  {Telford}, {Teplitz}, {Trump}, {van der Wel}, {Villforth}, {Wechsler},
  {Weiner}, {Wiklind}, {Wild}, {Wilson}, {Wuyts}, {Yan}, \& {Yun}}]{grogin11}
{Grogin}, N.~A., {Kocevski}, D.~D., {Faber}, S.~M., {et~al.} 2011, \apjs, 197,
  35

\bibitem[{{Harikane} {et~al.}(2018){Harikane}, {Ouchi}, {Ono}, {Saito},
  {Behroozi}, {More}, {Shimasaku}, {Toshikawa}, {Lin}, {Akiyama}, {Coupon},
  {Komiyama}, {Konno}, {Lin}, {Miyazaki}, {Nishizawa}, {Shibuya}, \&
  {Silverman}}]{harikane18}
{Harikane}, Y., {Ouchi}, M., {Ono}, Y., {et~al.} 2018, \pasj, 70, S11

\bibitem[{{Holwerda} {et~al.}(2015){Holwerda}, {Bouwens}, {Oesch}, {Smit},
  {Illingworth}, \& {Labbe}}]{holwerda15}
{Holwerda}, B.~W., {Bouwens}, R., {Oesch}, P., {et~al.} 2015, \apj, 808, 6

\bibitem[{{Holwerda} {et~al.}(2014){Holwerda}, {Trenti}, {Clarkson}, {Sahu},
  {Bradley}, {Stiavelli}, {Pirzkal}, {De Marchi}, {Andersen}, {Bouwens}, \&
  {Ryan}}]{holwerda14b}
{Holwerda}, B.~W., {Trenti}, M., {Clarkson}, W., {et~al.} 2014, \apj, 788, 77

\bibitem[{{Ilbert} {et~al.}(2009){Ilbert}, {Capak}, {Salvato}, {Aussel},
  {McCracken}, {Sanders}, {Scoville}, {Kartaltepe}, {Arnouts}, {Le Floc'h},
  {Mobasher}, {Taniguchi}, {Lamareille}, {Leauthaud}, {Sasaki}, {Thompson},
  {Zamojski}, {Zamorani}, {Bardelli}, {Bolzonella}, {Bongiorno}, {Brusa},
  {Caputi}, {Carollo}, {Contini}, {Cook}, {Coppa}, {Cucciati}, {de la Torre},
  {de Ravel}, {Franzetti}, {Garilli}, {Hasinger}, {Iovino}, {Kampczyk},
  {Kneib}, {Knobel}, {Kovac}, {Le Borgne}, {Le Brun}, {F{\`e}vre}, {Lilly},
  {Looper}, {Maier}, {Mainieri}, {Mellier}, {Mignoli}, {Murayama}, {Pell{\`o}},
  {Peng}, {P{\'e}rez-Montero}, {Renzini}, {Ricciardelli}, {Schiminovich},
  {Scodeggio}, {Shioya}, {Silverman}, {Surace}, {Tanaka}, {Tasca}, {Tresse},
  {Vergani}, \& {Zucca}}]{ilbert09}
{Ilbert}, O., {Capak}, P., {Salvato}, M., {et~al.} 2009, \apj, 690, 1236

\bibitem[{{Ishigaki} {et~al.}(2015){Ishigaki}, {Kawamata}, {Ouchi}, {Oguri},
  {Shimasaku}, \& {Ono}}]{ishigaki15}
{Ishigaki}, M., {Kawamata}, R., {Ouchi}, M., {et~al.} 2015, \apj, 799, 12

\bibitem[{{Ishigaki} {et~al.}(2018){Ishigaki}, {Kawamata}, {Ouchi}, {Oguri},
  {Shimasaku}, \& {Ono}}]{ishigaki18}
---. 2018, \apj, 854, 73

\bibitem[{{Ishigaki} {et~al.}(2016){Ishigaki}, {Ouchi}, \&
  {Harikane}}]{ishigaki16}
{Ishigaki}, M., {Ouchi}, M., \& {Harikane}, Y. 2016, \apj, 822, 5

\bibitem[{{Jaacks} {et~al.}(2012){Jaacks}, {Choi}, {Nagamine}, {Thompson}, \&
  {Varghese}}]{jaacks12}
{Jaacks}, J., {Choi}, J.-H., {Nagamine}, K., {Thompson}, R., \& {Varghese}, S.
  2012, \mnras, 420, 1606

\bibitem[{{Kalirai}(2018)}]{kalirai18}
{Kalirai}, J. 2018, Contemporary Physics, 59, 251

\bibitem[{{Kawamata} {et~al.}(2015){Kawamata}, {Ishigaki}, {Shimasaku},
  {Oguri}, \& {Ouchi}}]{kawamata15}
{Kawamata}, R., {Ishigaki}, M., {Shimasaku}, K., {Oguri}, M., \& {Ouchi}, M.
  2015, \apj, 804, 103

\bibitem[{{Kawamata} {et~al.}(2017){Kawamata}, {Ishigaki}, {Shimasaku},
  {Oguri}, {Ouchi}, \& {Tanigawa}}]{kawamata17}
{Kawamata}, R., {Ishigaki}, M., {Shimasaku}, K., {et~al.} 2017, ArXiv e-prints,
  arXiv:1710.07301

\bibitem[{{Kawamata} {et~al.}(2018){Kawamata}, {Ishigaki}, {Shimasaku},
  {Oguri}, {Ouchi}, \& {Tanigawa}}]{kawamata18}
---. 2018, \apj, 855, 4

\bibitem[{{Kelson} {et~al.}(2016){Kelson}, {Benson}, \& {Abramson}}]{kelson16}
{Kelson}, D.~D., {Benson}, A.~J., \& {Abramson}, L.~E. 2016, ArXiv e-prints,
  arXiv:1610.06566

\bibitem[{{Koekemoer} {et~al.}(2011){Koekemoer}, {Faber}, {Ferguson}, {Grogin},
  {Kocevski}, {Koo}, {Lai}, {Lotz}, {Lucas}, {McGrath}, {Ogaz}, {Rajan},
  {Riess}, {Rodney}, {Strolger}, {Casertano}, {Castellano}, {Dahlen},
  {Dickinson}, {Dolch}, {Fontana}, {Giavalisco}, {Grazian}, {Guo}, {Hathi},
  {Huang}, {van der Wel}, {Yan}, {Acquaviva}, {Alexander}, {Almaini}, {Ashby},
  {Barden}, {Bell}, {Bournaud}, {Brown}, {Caputi}, {Cassata}, {Challis},
  {Chary}, {Cheung}, {Cirasuolo}, {Conselice}, {Roshan Cooray}, {Croton},
  {Daddi}, {Dav{\'e}}, {de Mello}, {de Ravel}, {Dekel}, {Donley}, {Dunlop},
  {Dutton}, {Elbaz}, {Fazio}, {Filippenko}, {Finkelstein}, {Frazer}, {Gardner},
  {Garnavich}, {Gawiser}, {Gruetzbauch}, {Hartley}, {H{\"a}ussler},
  {Herrington}, {Hopkins}, {Huang}, {Jha}, {Johnson}, {Kartaltepe},
  {Khostovan}, {Kirshner}, {Lani}, {Lee}, {Li}, {Madau}, {McCarthy},
  {McIntosh}, {McLure}, {McPartland}, {Mobasher}, {Moreira}, {Mortlock},
  {Moustakas}, {Mozena}, {Nandra}, {Newman}, {Nielsen}, {Niemi}, {Noeske},
  {Papovich}, {Pentericci}, {Pope}, {Primack}, {Ravindranath}, {Reddy},
  {Renzini}, {Rix}, {Robaina}, {Rosario}, {Rosati}, {Salimbeni}, {Scarlata},
  {Siana}, {Simard}, {Smidt}, {Snyder}, {Somerville}, {Spinrad}, {Straughn},
  {Telford}, {Teplitz}, {Trump}, {Vargas}, {Villforth}, {Wagner}, {Wandro},
  {Wechsler}, {Weiner}, {Wiklind}, {Wild}, {Wilson}, {Wuyts}, \&
  {Yun}}]{koekemoer11}
{Koekemoer}, A.~M., {Faber}, S.~M., {Ferguson}, H.~C., {et~al.} 2011, \apjs,
  197, 36

\bibitem[{{Konno} {et~al.}(2014){Konno}, {Ouchi}, {Ono}, {Shimasaku},
  {Shibuya}, {Furusawa}, {Nakajima}, {Naito}, {Momose}, {Yuma}, \&
  {Iye}}]{konno14}
{Konno}, A., {Ouchi}, M., {Ono}, Y., {et~al.} 2014, \apj, 797, 16

\bibitem[{{Kriek} {et~al.}(2009){Kriek}, {van Dokkum}, {Labb{\'e}}, {Franx},
  {Illingworth}, {Marchesini}, \& {Quadri}}]{kriek09}
{Kriek}, M., {van Dokkum}, P.~G., {Labb{\'e}}, I., {et~al.} 2009, \apj, 700,
  221

\bibitem[{{Livermore} {et~al.}(2017){Livermore}, {Finkelstein}, \&
  {Lotz}}]{livermore17}
{Livermore}, R.~C., {Finkelstein}, S.~L., \& {Lotz}, J.~M. 2017, \apj, 835, 113

\bibitem[{{Livermore} {et~al.}(2018){Livermore}, {Trenti}, {Bradley},
  {Bernard}, {Holwerda}, {Mason}, \& {Treu}}]{livermore18}
{Livermore}, R.~C., {Trenti}, M., {Bradley}, L.~D., {et~al.} 2018, \apjl, 861,
  L17

\bibitem[{{Madau} \& {Dickinson}(2014)}]{madau14}
{Madau}, P., \& {Dickinson}, M. 2014, \araa, 52, 415

\bibitem[{{Maraston}(2005)}]{maraston05}
{Maraston}, C. 2005, \mnras, 362, 799

\bibitem[{{Mason} {et~al.}(2015{\natexlab{a}}){Mason}, {Trenti}, \&
  {Treu}}]{mason15b}
{Mason}, C.~A., {Trenti}, M., \& {Treu}, T. 2015{\natexlab{a}}, \apj, 813, 21

\bibitem[{{Mason} {et~al.}(2018{\natexlab{a}}){Mason}, {Treu}, {Dijkstra},
  {Mesinger}, {Trenti}, {Pentericci}, {de Barros}, \& {Vanzella}}]{mason18}
{Mason}, C.~A., {Treu}, T., {Dijkstra}, M., {et~al.} 2018{\natexlab{a}}, \apj,
  856, 2

\bibitem[{{Mason} {et~al.}(2015{\natexlab{b}}){Mason}, {Treu}, {Schmidt},
  {Collett}, {Trenti}, {Marshall}, {Barone-Nugent}, {Bradley}, {Stiavelli}, \&
  {Wyithe}}]{mason15}
{Mason}, C.~A., {Treu}, T., {Schmidt}, K.~B., {et~al.} 2015{\natexlab{b}},
  \apj, 805, 79

\bibitem[{{Mason} {et~al.}(2018{\natexlab{b}}){Mason}, {Treu}, {de Barros},
  {Dijkstra}, {Fontana}, {Mesinger}, {Pentericci}, {Trenti}, \&
  {Vanzella}}]{mason18b}
{Mason}, C.~A., {Treu}, T., {de Barros}, S., {et~al.} 2018{\natexlab{b}},
  \apjl, 857, L11

\bibitem[{{McLeod} {et~al.}(2016){McLeod}, {McLure}, \& {Dunlop}}]{mcleod16}
{McLeod}, D.~J., {McLure}, R.~J., \& {Dunlop}, J.~S. 2016, \mnras, 459, 3812

\bibitem[{{McLure} {et~al.}(2013){McLure}, {Dunlop}, {Bowler}, {Curtis-Lake},
  {Schenker}, {Ellis}, {Robertson}, {Koekemoer}, {Rogers}, {Ono}, {Ouchi},
  {Charlot}, {Wild}, {Stark}, {Furlanetto}, {Cirasuolo}, \&
  {Targett}}]{mclure13}
{McLure}, R.~J., {Dunlop}, J.~S., {Bowler}, R.~A.~A., {et~al.} 2013, \mnras,
  432, 2696

\bibitem[{{Merlin} {et~al.}(2016){Merlin}, {Bourne}, {Castellano}, {Ferguson},
  {Wang}, {Derriere}, {Dunlop}, {Elbaz}, \& {Fontana}}]{merlin16}
{Merlin}, E., {Bourne}, N., {Castellano}, M., {et~al.} 2016, \aap, 595, A97

\bibitem[{{Mu{\~n}oz} \& {Loeb}(2008)}]{munoz08}
{Mu{\~n}oz}, J.~A., \& {Loeb}, A. 2008, \mnras, 385, 2175

\bibitem[{{Noeske} {et~al.}(2012){Noeske}, {Baggett}, {Bushouse}, {Petro},
  {Gilliland}, \& {Khozurina-Platais}}]{noeske12}
{Noeske}, K., {Baggett}, S., {Bushouse}, H., {et~al.} 2012, {WFC3 UVIS Charge
  Transfer Eciency October 2009 to October 2011}, Tech. rep.

\bibitem[{{Oesch} {et~al.}(2018){Oesch}, {Bouwens}, {Illingworth}, {Labb{\'e}},
  \& {Stefanon}}]{oesch18}
{Oesch}, P.~A., {Bouwens}, R.~J., {Illingworth}, G.~D., {Labb{\'e}}, I., \&
  {Stefanon}, M. 2018, \apj, 855, 105

\bibitem[{{Oesch} {et~al.}(2007){Oesch}, {Stiavelli}, {Carollo}, {Bergeron},
  {Koekemoer}, {Lucas}, {Pavlovsky}, {Trenti}, {Lilly}, {Beckwith}, {Dahlen},
  {Ferguson}, {Gardner}, {Lacey}, {Mobasher}, {Panagia}, \& {Rix}}]{oesch07}
{Oesch}, P.~A., {Stiavelli}, M., {Carollo}, C.~M., {et~al.} 2007, \apj, 671,
  1212

\bibitem[{{Oesch} {et~al.}(2010){Oesch}, {Bouwens}, {Carollo}, {Illingworth},
  {Trenti}, {Stiavelli}, {Magee}, {Labb{\'e}}, \& {Franx}}]{oesch10}
{Oesch}, P.~A., {Bouwens}, R.~J., {Carollo}, C.~M., {et~al.} 2010, \apjl, 709,
  L21

\bibitem[{{Oesch} {et~al.}(2012){Oesch}, {Bouwens}, {Illingworth}, {Labb{\'e}},
  {Trenti}, {Gonzalez}, {Carollo}, {Franx}, {van Dokkum}, \& {Magee}}]{oesch12}
{Oesch}, P.~A., {Bouwens}, R.~J., {Illingworth}, G.~D., {et~al.} 2012, \apj,
  745, 110

\bibitem[{{Oesch} {et~al.}(2013){Oesch}, {Bouwens}, {Illingworth}, {Labb{\'e}},
  {Franx}, {van Dokkum}, {Trenti}, {Stiavelli}, {Gonzalez}, \&
  {Magee}}]{oesch13}
---. 2013, \apj, 773, 75

\bibitem[{{Oesch} {et~al.}(2016){Oesch}, {Brammer}, {van Dokkum},
  {Illingworth}, {Bouwens}, {Labb{\'e}}, {Franx}, {Momcheva}, {Ashby}, {Fazio},
  {Gonzalez}, {Holden}, {Magee}, {Skelton}, {Smit}, {Spitler}, {Trenti}, \&
  {Willner}}]{oesch16}
{Oesch}, P.~A., {Brammer}, G., {van Dokkum}, P.~G., {et~al.} 2016, \apj, 819,
  129

\bibitem[{{Oke} \& {Gunn}(1983)}]{oke83}
{Oke}, J.~B., \& {Gunn}, J.~E. 1983, \apj, 266, 713

\bibitem[{{Ono} {et~al.}(2013){Ono}, {Ouchi}, {Curtis-Lake}, {Schenker},
  {Ellis}, {McLure}, {Dunlop}, {Robertson}, {Koekemoer}, {Bowler}, {Rogers},
  {Schneider}, {Charlot}, {Stark}, {Shimasaku}, {Furlanetto}, \&
  {Cirasuolo}}]{ono13}
{Ono}, Y., {Ouchi}, M., {Curtis-Lake}, E., {et~al.} 2013, \apj, 777, 155

\bibitem[{{Ono} {et~al.}(2018){Ono}, {Ouchi}, {Harikane}, {Toshikawa}, {Rauch},
  {Yuma}, {Sawicki}, {Shibuya}, {Shimasaku}, {Oguri}, {Willott}, {Akhlaghi},
  {Akiyama}, {Coupon}, {Kashikawa}, {Komiyama}, {Konno}, {Lin}, {Matsuoka},
  {Miyazaki}, {Nagao}, {Nakajima}, {Silverman}, {Tanaka}, {Taniguchi}, \&
  {Wang}}]{ono18}
{Ono}, Y., {Ouchi}, M., {Harikane}, Y., {et~al.} 2018, \pasj, 70, S10

\bibitem[{{Ouchi} {et~al.}(2010){Ouchi}, {Shimasaku}, {Furusawa}, {Saito},
  {Yoshida}, {Akiyama}, {Ono}, {Yamada}, {Ota}, {Kashikawa}, {Iye}, {Kodama},
  {Okamura}, {Simpson}, \& {Yoshida}}]{ouchi10}
{Ouchi}, M., {Shimasaku}, K., {Furusawa}, H., {et~al.} 2010, \apj, 723, 869

\bibitem[{{Peng} {et~al.}(2010){Peng}, {Lilly}, {Kova{\v c}}, {Bolzonella},
  {Pozzetti}, {Renzini}, {Zamorani}, {Ilbert}, {Knobel}, {Iovino}, {Maier},
  {Cucciati}, {Tasca}, {Carollo}, {Silverman}, {Kampczyk}, {de Ravel},
  {Sanders}, {Scoville}, {Contini}, {Mainieri}, {Scodeggio}, {Kneib}, {Le
  F{\`e}vre}, {Bardelli}, {Bongiorno}, {Caputi}, {Coppa}, {de la Torre},
  {Franzetti}, {Garilli}, {Lamareille}, {Le Borgne}, {Le Brun}, {Mignoli},
  {Perez Montero}, {Pello}, {Ricciardelli}, {Tanaka}, {Tresse}, {Vergani},
  {Welikala}, {Zucca}, {Oesch}, {Abbas}, {Barnes}, {Bordoloi}, {Bottini},
  {Cappi}, {Cassata}, {Cimatti}, {Fumana}, {Hasinger}, {Koekemoer},
  {Leauthaud}, {Maccagni}, {Marinoni}, {McCracken}, {Memeo}, {Meneux}, {Nair},
  {Porciani}, {Presotto}, \& {Scaramella}}]{peng10}
{Peng}, Y.-j., {Lilly}, S.~J., {Kova{\v c}}, K., {et~al.} 2010, \apj, 721, 193

\bibitem[{{Pirzkal} {et~al.}(2013){Pirzkal}, {Rothberg}, {Ryan}, {Coe},
  {Malhotra}, {Rhoads}, \& {Noeske}}]{pirzkal13}
{Pirzkal}, N., {Rothberg}, B., {Ryan}, R., {et~al.} 2013, \apj, 775, 11

\bibitem[{{Planck Collaboration} {et~al.}(2014){Planck Collaboration}, {Ade},
  {Aghanim}, {Armitage-Caplan}, {Arnaud}, {Ashdown}, {Atrio-Barandela},
  {Aumont}, {Baccigalupi}, {Banday}, \& et~al.}]{pc1316}
{Planck Collaboration}, {Ade}, P.~A.~R., {Aghanim}, N., {et~al.} 2014, \aap,
  571, A16

\bibitem[{{Ren} {et~al.}(2018){Ren}, {Trenti}, \& {Mutch}}]{ren18}
{Ren}, K., {Trenti}, M., \& {Mutch}, S.~J. 2018, \apj, 856, 81

\bibitem[{{Robertson} {et~al.}(2010){Robertson}, {Ellis}, {Dunlop}, {McLure},
  \& {Stark}}]{robertson10}
{Robertson}, B.~E., {Ellis}, R.~S., {Dunlop}, J.~S., {McLure}, R.~J., \&
  {Stark}, D.~P. 2010, \nat, 468, 49

\bibitem[{{Salmon} {et~al.}(2017){Salmon}, {Coe}, {Bradley}, {Bouwens},
  {Bradac}, {Huang}, {Oesch}, {Stark}, {Sharon}, {Trenti}, {Avila}, {Ogaz},
  {Andrade-Santos}, {Carrasco}, {Cerny}, {Dawson}, {Frye}, {Hoag}, {Johnson},
  {Jones}, {Lam}, {Lovisari}, {Mainali}, {Past}, {Paterno-Mahler}, {Peterson},
  {Reiss}, {Rodney}, {Ryan}, {Sendra-Server}, {Strolger}, {Umetsu}, {Vulcani},
  \& {Zitrin}}]{salmon17}
{Salmon}, B., {Coe}, D., {Bradley}, L., {et~al.} 2017, ArXiv e-prints,
  arXiv:1710.08930

\bibitem[{{Salmon} {et~al.}(2018){Salmon}, {Coe}, {Bradley}, {Brada{\v c}},
  {Huang}, {Strait}, {Oesch}, {Paterno-Mahler}, {Zitrin}, {Acebron}, {Cibirka},
  {Kikuchihara}, {Oguri}, {Brammer}, {Sharon}, {Trenti}, {Avila}, {Ogaz},
  {Andrade-Santos}, {Carrasco}, {Cerny}, {Dawson}, {Frye}, {Hoag}, {Jones},
  {Mainali}, {Ouchi}, {Rodney}, {Stark}, \& {Umetsu}}]{salmon18}
---. 2018, ArXiv e-prints, arXiv:1801.03103

\bibitem[{{Salpeter}(1955)}]{salpeter55}
{Salpeter}, E.~E. 1955, \apj, 121, 161

\bibitem[{{Salvato} {et~al.}(2018){Salvato}, {Ilbert}, \& {Hoyle}}]{salvato18}
{Salvato}, M., {Ilbert}, O., \& {Hoyle}, B. 2018, Nature Astronomy,
  arXiv:1805.12574

\bibitem[{{Schenker} {et~al.}(2013){Schenker}, {Robertson}, {Ellis}, {Ono},
  {McLure}, {Dunlop}, {Koekemoer}, {Bowler}, {Ouchi}, {Curtis-Lake}, {Rogers},
  {Schneider}, {Charlot}, {Stark}, {Furlanetto}, \& {Cirasuolo}}]{schenker13}
{Schenker}, M.~A., {Robertson}, B.~E., {Ellis}, R.~S., {et~al.} 2013, \apj,
  768, 196

\bibitem[{{Schlafly} \& {Finkbeiner}(2011)}]{schlafly11}
{Schlafly}, E.~F., \& {Finkbeiner}, D.~P. 2011, \apj, 737, 103

\bibitem[{{Schlegel} {et~al.}(1998){Schlegel}, {Finkbeiner}, \&
  {Davis}}]{schlegel98}
{Schlegel}, D.~J., {Finkbeiner}, D.~P., \& {Davis}, M. 1998, \apj, 500, 525

\bibitem[{{Schmidt} {et~al.}(2014){Schmidt}, {Treu}, {Trenti}, {Bradley},
  {Kelly}, {Oesch}, {Holwerda}, {Shull}, \& {Stiavelli}}]{schmidt14b}
{Schmidt}, K.~B., {Treu}, T., {Trenti}, M., {et~al.} 2014, \apj, 786, 57

\bibitem[{{Sersic}(1968)}]{sersic68}
{Sersic}, J.~L. 1968, {Atlas de galaxias australes}

\bibitem[{{Shibuya} {et~al.}(2015){Shibuya}, {Ouchi}, \&
  {Harikane}}]{shibuya15}
{Shibuya}, T., {Ouchi}, M., \& {Harikane}, Y. 2015, \apjs, 219, 15

\bibitem[{{Shull} {et~al.}(2012){Shull}, {Harness}, {Trenti}, \&
  {Smith}}]{shull12}
{Shull}, J.~M., {Harness}, A., {Trenti}, M., \& {Smith}, B.~D. 2012, \apj, 747,
  100

\bibitem[{{Sobral} {et~al.}(2015){Sobral}, {Matthee}, {Darvish}, {Schaerer},
  {Mobasher}, {R{\"o}ttgering}, {Santos}, \& {Hemmati}}]{sobral15}
{Sobral}, D., {Matthee}, J., {Darvish}, B., {et~al.} 2015, \apj, 808, 139

\bibitem[{{Somerville} {et~al.}(2004){Somerville}, {Lee}, {Ferguson},
  {Gardner}, {Moustakas}, \& {Giavalisco}}]{somerville04}
{Somerville}, R.~S., {Lee}, K., {Ferguson}, H.~C., {et~al.} 2004, \apjl, 600,
  L171

\bibitem[{{Spergel} {et~al.}(2015){Spergel}, {Gehrels}, {Baltay}, {Bennett},
  {Breckinridge}, {Donahue}, {Dressler}, {Gaudi}, {Greene}, {Guyon}, {Hirata},
  {Kalirai}, {Kasdin}, {Macintosh}, {Moos}, {Perlmutter}, {Postman},
  {Rauscher}, {Rhodes}, {Wang}, {Weinberg}, {Benford}, {Hudson}, {Jeong},
  {Mellier}, {Traub}, {Yamada}, {Capak}, {Colbert}, {Masters}, {Penny},
  {Savransky}, {Stern}, {Zimmerman}, {Barry}, {Bartusek}, {Carpenter}, {Cheng},
  {Content}, {Dekens}, {Demers}, {Grady}, {Jackson}, {Kuan}, {Kruk}, {Melton},
  {Nemati}, {Parvin}, {Poberezhskiy}, {Peddie}, {Ruffa}, {Wallace}, {Whipple},
  {Wollack}, \& {Zhao}}]{spergel15}
{Spergel}, D., {Gehrels}, N., {Baltay}, C., {et~al.} 2015, ArXiv e-prints,
  arXiv:1503.03757

\bibitem[{{Stefanon} {et~al.}(2017){Stefanon}, {Labb{\'e}}, {Bouwens},
  {Brammer}, {Oesch}, {Franx}, {Fynbo}, {Milvang-Jensen}, {Muzzin},
  {Illingworth}, {Le F{\`e}vre}, {Caputi}, {Holwerda}, {McCracken}, {Smit}, \&
  {Magee}}]{stefanon17}
{Stefanon}, M., {Labb{\'e}}, I., {Bouwens}, R.~J., {et~al.} 2017, \apj, 851, 43

\bibitem[{{Stiavelli}(2009)}]{stiavelli09}
{Stiavelli}, M. 2009, {From First Light to Reionization: The End of the Dark
  Ages}

\bibitem[{{Tacchella} {et~al.}(2013){Tacchella}, {Trenti}, \&
  {Carollo}}]{tacchella13}
{Tacchella}, S., {Trenti}, M., \& {Carollo}, C.~M. 2013, \apjl, 768, L37

\bibitem[{{Totani} {et~al.}(2006){Totani}, {Kawai}, {Kosugi}, {Aoki}, {Yamada},
  {Iye}, {Ohta}, \& {Hattori}}]{totani06}
{Totani}, T., {Kawai}, N., {Kosugi}, G., {et~al.} 2006, \pasj, 58, 485

\bibitem[{{Trac} {et~al.}(2015){Trac}, {Cen}, \& {Mansfield}}]{trac15}
{Trac}, H., {Cen}, R., \& {Mansfield}, P. 2015, \apj, 813, 54

\bibitem[{{Trenti} {et~al.}(2012{\natexlab{a}}){Trenti}, {Perna}, {Levesque},
  {Shull}, \& {Stocke}}]{trenti12b}
{Trenti}, M., {Perna}, R., {Levesque}, E.~M., {Shull}, J.~M., \& {Stocke},
  J.~T. 2012{\natexlab{a}}, \apjl, 749, L38

\bibitem[{{Trenti} \& {Stiavelli}(2008)}]{trenti08}
{Trenti}, M., \& {Stiavelli}, M. 2008, \apj, 676, 767

\bibitem[{{Trenti} {et~al.}(2010){Trenti}, {Stiavelli}, {Bouwens}, {Oesch},
  {Shull}, {Illingworth}, {Bradley}, \& {Carollo}}]{trenti10}
{Trenti}, M., {Stiavelli}, M., {Bouwens}, R.~J., {et~al.} 2010, \apjl, 714,
  L202

\bibitem[{{Trenti} {et~al.}(2011){Trenti}, {Bradley}, {Stiavelli}, {Oesch},
  {Treu}, {Bouwens}, {Shull}, {MacKenty}, {Carollo}, \&
  {Illingworth}}]{trenti11}
{Trenti}, M., {Bradley}, L.~D., {Stiavelli}, M., {et~al.} 2011, \apjl, 727, L39

\bibitem[{{Trenti} {et~al.}(2012{\natexlab{b}}){Trenti}, {Bradley},
  {Stiavelli}, {Shull}, {Oesch}, {Bouwens}, {Mu{\~n}oz}, {Romano-Diaz}, {Treu},
  {Shlosman}, \& {Carollo}}]{trenti12}
---. 2012{\natexlab{b}}, \apj, 746, 55

\bibitem[{{Treu} {et~al.}(2013){Treu}, {Schmidt}, {Trenti}, {Bradley}, \&
  {Stiavelli}}]{treu13}
{Treu}, T., {Schmidt}, K.~B., {Trenti}, M., {Bradley}, L.~D., \& {Stiavelli},
  M. 2013, \apjl, 775, L29

\bibitem[{{Treu} {et~al.}(2012){Treu}, {Trenti}, {Stiavelli}, {Auger}, \&
  {Bradley}}]{treu12}
{Treu}, T., {Trenti}, M., {Stiavelli}, M., {Auger}, M.~W., \& {Bradley}, L.~D.
  2012, \apj, 747, 27

\bibitem[{{van Dokkum} {et~al.}(2001){van Dokkum}, {Franx}, {Kelson}, \&
  {Illingworth}}]{vandokkum01}
{van Dokkum}, P.~G., {Franx}, M., {Kelson}, D.~D., \& {Illingworth}, G.~D.
  2001, \apjl, 553, L39

\bibitem[{{Vulcani} {et~al.}(2017){Vulcani}, {Trenti}, {Calvi}, {Bouwens},
  {Oesch}, {Stiavelli}, \& {Franx}}]{vulcani17}
{Vulcani}, B., {Trenti}, M., {Calvi}, V., {et~al.} 2017, \apj, 836, 239

\bibitem[{{Wilkins} {et~al.}(2016){Wilkins}, {Bouwens}, {Oesch}, {Labb{\'e}},
  {Sargent}, {Caruana}, {Wardlow}, \& {Clay}}]{wilkins16}
{Wilkins}, S.~M., {Bouwens}, R.~J., {Oesch}, P.~A., {et~al.} 2016, \mnras, 455,
  659

\bibitem[{{Williams} {et~al.}(2018){Williams}, {Curtis-Lake}, {Hainline},
  {Chevallard}, {Robertson}, {Charlot}, {Endsley}, {Stark}, {Willmer},
  {Alberts}, {Amorin}, {Arribas}, {Baum}, {Bunker}, {Carniani}, {Crandall},
  {Egami}, {Eisenstein}, {Ferruit}, {Husemann}, {Maseda}, {Maiolino}, {Rawle},
  {Rieke}, {Smit}, {Tacchella}, \& {Willott}}]{williams18}
{Williams}, C.~C., {Curtis-Lake}, E., {Hainline}, K.~N., {et~al.} 2018, \apjs,
  236, 33

\bibitem[{{Woo} {et~al.}(2013){Woo}, {Dekel}, {Faber}, {Noeske}, {Koo},
  {Gerke}, {Cooper}, {Salim}, {Dutton}, {Newman}, {Weiner}, {Bundy}, {Willmer},
  {Davis}, \& {Yan}}]{woo13}
{Woo}, J., {Dekel}, A., {Faber}, S.~M., {et~al.} 2013, \mnras, 428, 3306

\bibitem[{{Yung} {et~al.}(2018){Yung}, {Somerville}, {Finkelstein}, {Popping},
  \& {Dav{\'e}}}]{yung18}
{Yung}, L.~Y.~A., {Somerville}, R.~S., {Finkelstein}, S.~L., {Popping}, G., \&
  {Dav{\'e}}, R. 2018, ArXiv e-prints, arXiv:1803.09761

\bibitem[{{Zitrin} {et~al.}(2015){Zitrin}, {Labb{\'e}}, {Belli}, {Bouwens},
  {Ellis}, {Roberts-Borsani}, {Stark}, {Oesch}, \& {Smit}}]{zitrin15}
{Zitrin}, A., {Labb{\'e}}, I., {Belli}, S., {et~al.} 2015, \apjl, 810, L12

\end{thebibliography}
\end{document}